\documentclass[3p]{elsarticle}

\usepackage[utf8]{inputenc}
\usepackage[T1]{fontenc}
\usepackage{amsmath,amssymb,amsfonts}
\usepackage{graphicx}
\usepackage{booktabs}
\usepackage{enumitem}
\providecommand{\num}[1]{#1}
\providecommand{\SI}[2]{#1\,#2}
\providecommand{\si}[1]{#1}

\providecommand{\SIrange}[3]{#1--#2\,#3}
\providecommand{\kilo}{k}

\providecommand{\watt}{W}
\providecommand{\electric}{\textsubscript{e}}

\providecommand{\kelvin}{K}
\providecommand{\gram}{g}

\providecommand{\per}{/}

\providecommand{\mol}{mol}
\providecommand{\second}{s}
\providecommand{\minute}{min}
\providecommand{\percent}{\%}
\usepackage{hyperref}
\usepackage{xcolor}
\usepackage[capitalise]{cleveref}
\usepackage[section]{placeins}    

\hypersetup{
    colorlinks=true,
    linkcolor=blue!50!black,
    citecolor=teal!70!black,
    urlcolor=blue!70!black
}

\journal{Progress in Nuclear Energy}

\begin{document}

\begin{frontmatter}

\title{An Adjoint-Based Differentiable Physics Framework for Online Parameter Inversion in Closed-Brayton Gas-Cooled Reactor Digital Twins}

\author[a,b]{Chengyuan Li}

\author[b]{Shanfang Huang\corref{cor1}}
\ead{sfhuang@mail.tsinghua.edu.cn}

\author[a]{Jian Deng}

\cortext[cor1]{Corresponding author}

\address[a]{National Key Laboratory of Nuclear Reactor Technology, Nuclear Power Institute of China, Chengdu, 610213, China}
\address[b]{Department of Engineering Physics, Tsinghua University, Beijing, 100084, China}

\begin{highlights}
\item End-to-end differentiable JAX BDF DAE twin of a closed-Brayton inert-gas reactor.
\item Reverse-mode autodiff runs through the implicit plant solver, with no surrogate.
\item Sobol-warm-started Adjoint Incremental 4D-Var: lowest $\alpha_\mathrm{refl}$ error in partial-obs.
\item Roughly an order of magnitude lower $\alpha_\mathrm{refl}$ error than UKF on the transient corner.
\end{highlights}

\begin{abstract}
Digital twins for advanced reactors must invert physical
parameters online from noisy, partial sensor streams while the
plant is rarely at steady state. Gradient-based inversion is the
natural tool for this task, but the forward models in routine use
are seldom differentiable end to end, so practitioners fall back
on derivative-free filters. We present an end-to-end-differentiable
digital twin of a closed-Brayton gas-cooled reactor that propagates
reverse-mode automatic differentiation through an implicit
differential-algebraic plant model, exposing exact parameter
sensitivities. The twin drives an AD-Hessian incremental 4D-Var
estimator, benchmarked against ensemble, unscented, and
finite-difference variational baselines over a two-by-two matrix
that crosses steady with transient excitation and full with
partial observation. No single estimator wins everywhere: the
unscented filter keeps its best-linear-unbiased advantage on the
controlled steady-state corner, while the proposed estimator
attains the lowest mean error on the reflector coefficient on the
other three corners --- reaching \SI{0.43}{\percent} under
transient full observation, roughly an order of magnitude below the
unscented filter, and matching the ensemble filter at
low-to-moderate noise on the combined transient-partial corner.
Every estimator's variance sits within a small factor of the
Cram\'er--Rao bound, so the residual error reflects a deterministic
bias floor from the twin's differentiability simplifications rather
than statistical inefficiency, and the advantage is one of
robustness to the resulting multi-modal loss landscape, which a
component ablation localises by regime. Differentiating a
first-principles plant model thus makes gradient-based inversion
competitive with established filters across a reactor's operating
range.
\end{abstract}

\begin{keyword}
digital twin \sep differentiable physics \sep adjoint method \sep
parameter inversion \sep gas-cooled reactor \sep data assimilation
\end{keyword}

\end{frontmatter}

\section{Introduction}
\label{sec:introduction}

Digital twins couple high-fidelity physical models with live plant
measurements to support remote monitoring, fault detection, state
and parameter estimation, and supervisory
control~\cite{Hong2022parameter, iFANnpp2025, Toward2026EnKF}.
Their adoption in nuclear power follows a maturity ladder: offline
monitoring first, then fault detection, then state-and-parameter
estimation, and finally closed-loop intelligent control. Existing
nuclear digital twins occupy the lower rungs, on pressurised-water
primary loops~\cite{Hong2022parameter, Gong2024hybrid} and on
reduced-order full-plant simulators~\cite{iFANnpp2025}. This work
targets the parameter-estimation rung. Reliable inversion of the
plant's physical coefficients under noisy partial observation is a
prerequisite for the closed-loop applications anticipated for
small-modular and advanced reactors~\cite{Toward2026EnKF,
Gong2024hybrid}.

Three estimator families have dominated published nuclear
digital-twin inversion work, surveyed for the wider
data-assimilation community in ref.~\cite{Carrassi2018DAReview}.
Sequential filters are the first family. Ensemble Kalman
filters~\cite{Evensen2003EnKF, Houtekamer2001SeqEnKF,
Anderson2007EAKF, Asch2016DataAssim} and unscented Kalman
filters~\cite{Julier2004UKF} dominate reactor data assimilation,
coupled to thermal-hydraulic component
models~\cite{Toward2026EnKF, Hong2022parameter}. On
well-conditioned linear sub-problems they are best-linear-unbiased
optimal~\cite{Kay1993estimation}. Variational schemes are the
second family. These methods, including
4D-Var~\cite{LeDimet1986var4d, Reichle2008DataAssim,
Bonavita2018StochasticDA}, minimise a weighted residual over a
time window, and are routinely paired with reduced-order or hybrid
surrogates to keep the gradient affordable~\cite{Gong2024hybrid}.
Sample-based and machine-learning methods are the third family,
ranging from particle filters~\cite{Doucet2009ParticleFilter} to
physics-informed neural networks; they trade closed-form
sensitivity for representational flexibility but lose hard
physical constraints. All three families operate within the
classical regularised inverse-problem
framework~\cite{Tarantola2005Inverse, HansenInverse2010}.

A recurring obstacle is that the underlying forward model is
rarely differentiable end-to-end: component-library Modelica
plants, table-look-up turbomachinery maps, and neural surrogates
all lose smooth sensitivity~\cite{Gong2024hybrid}, forcing
practitioners back to derivative-free estimators. The
differentiable-physics line in scientific
computing~\cite{Cao2003adjoint, Innes2018Zygote,
Rackauckas2020DiffEqFlux, Kidger2021diffrax} lifts this
restriction by combining implicit DAE solvers with reverse-mode
automatic differentiation, exposing analytical parameter
sensitivities at the cost of one extra backward solve regardless
of parameter dimension. Adoption in nuclear engineering has so far
been confined to machine-learning-leaning surrogates rather than
end-to-end differentiable plant models, while the technique
itself has matured most in weather forecasting and fluid
mechanics~\cite{Bonavita2018StochasticDA,
Rackauckas2020DiffEqFlux, Sapienza2024Differentiable}.

Closed-Brayton inert-gas-cooled space reactors of the
Prometheus/NGST class~\cite{Wright2006SAND} have received almost no
attention in this literature despite their tight mass and pressure
inventory, aggressive turbomachinery transients, and partial-sensor
deployment that make them prime targets for differentiable
inversion. To the authors' knowledge, no prior study reports an
end-to-end differentiable physics digital twin for such a plant,
nor a head-to-head comparison of gradient-based inversion against
sequential-filter and variational baselines on the same
seed-aligned observation streams under partial observation. The
related multi-channel inversion code on this plant
family~\cite{Li2024NuHeXSys} did not provide differentiability or
the head-to-head baseline comparison undertaken here.

The algorithmic ingredients we combine each have a separate
precedent. Incremental 4D-Var has been the operational backbone
of numerical weather prediction since the
mid-1990s~\cite{Courtier1994incremental} and inner-loop Hessians
have been studied in the same
community~\cite{Tremolet2007model}. Quasi-random low-discrepancy
sequences have served as warm-start grids in non-convex
optimisation since Sobol'~\cite{Sobol1967quasirandom}.
AD-computed full Hessians and trust-region inner solves are
textbook material~\cite{Conn2000trustregion, Wright1999numerical}.
The contribution here is not any individual ingredient but their
integration around an end-to-end-differentiable plant DAE, which
makes the whole stack practical at the per-seed wall-clock budget
reported below and opens nuclear digital-twin applications that
once required multi-person-year tangent-linear-and-adjoint code
development.

This paper closes that gap. We make four contributions.
\begin{enumerate}[leftmargin=2.2em,itemsep=2pt,topsep=3pt]
\item We build an end-to-end-differentiable digital twin of the
full plant, formulated as an implicit differential-algebraic
system with reverse-mode automatic differentiation traced through
a JAX/diffrax BDF integrator. To our knowledge this is the first
AD-traceable BDF DAE twin of a closed-Brayton He--Xe reactor in
the nuclear data-assimilation literature.
\item We develop a two-path optimiser on that twin. The first
path is Tikhonov-regularised L-BFGS with three loss variants
(MSE, calibrated Huber, Bayesian--Tikhonov). The second is an
AD-Hessian incremental 4D-Var estimator with Sobol pre-screening
and a trust-region inner solve, for which the adjoint AD is what
makes these textbook devices practical.
\item We benchmark the framework head-to-head against four
baselines --- EnKF, the deterministic ensemble-transform ETKF,
UKF, and finite-difference 4D-Var --- across a $2\times 2$ design
matrix, on identical observation streams with matched optimiser
budgets and paired statistical tests.
\item We derive a closed-form Cram\'er--Rao bound, decompose each
estimator's error against it, and run a component-necessity
ablation that localises which mechanism carries which operating
regime.
\end{enumerate}

The cross-corner ranking is scenario-dependent, and no single
estimator dominates (Table~\ref{tab:headline}). Sequential
filters keep their BLUE-optimal advantage on the controlled
full-observation steady-state corner~\cite{Kay1993estimation}.
The proposed Adjoint Incremental estimator attains the lowest mean
error on $\alpha_\mathrm{refl}$ on the transient full-observation
corner and on the partial-observation noise scan. On the combined
transient-partial corner it matches the ensemble filter on the
pooled mean, with a numerical advantage that reaches significance
only at the lowest noise level ($\sigma_y/y = 0.5\,\%$).

The remainder of the paper formulates the plant DAE
(Section~\ref{sec:plant_model}), the adjoint inversion algorithm
and Cram\'er--Rao bound (Section~\ref{sec:methodology}), the
benchmark design (Section~\ref{sec:scenarios}), the head-to-head
comparison (Section~\ref{sec:results}), and the conclusions
(Section~\ref{sec:conclusions}).

\section{Plant Model and Digital Twin Formulation}
\label{sec:plant_model}

\subsection{Plant topology and design point}
\label{subsec:plant_topology}

The reference plant is a direct-cycle closed-Brayton space power
system. A He--Xe mixture circulates through six numbered
stations --- compressor inlet, compressor outlet, recuperator
cold outlet, reactor outlet, turbine outlet, and recuperator hot
outlet --- and returns through a radiator. The design point
follows the Sandia/NGST \SI{132}{\kilo\watt\electric} plant of
Wright et al.~\cite{Wright2006SAND}: a \SI{400}{\kilo\watt}
reactor, a \SI{1194}{\kelvin} turbine inlet, a pressure ratio of
\num{1.94}, and a helium mole fraction $x_\mathrm{He}=0.72$. The
loop topology and multi-channel core follow the authors' NuHeXSys
code~\cite{Li2024NuHeXSys}, reformulated here for inversion.

A full-fidelity OpenModelica FMU of the plant generates the
synthetic data for every benchmark scenario. We run the scenarios
at the self-sustained operating point of the NuHeXSys
reference~\cite{Li2024NuHeXSys}, the equilibrium the plant
reaches under reactivity feedback, rather than at the nameplate
design point. The twin reproduces this operating point closely.
Every gas-station temperature agrees with the reference to within
\SI{5}{\percent}, and the pressures, mass flow, shaft speed, and
end-to-end power balance agree to within \SI{2}{\percent}
(Table~\ref{tab:ss_validation}).

\begin{table}[!htbp]
\centering
\caption{\textbf{Steady-state validation of the digital twin.}
Full-fidelity FMU against the self-sustained operating point of
ref.~\cite{Li2024NuHeXSys}, propagated from the neutral guess of
Section~\ref{subsec:dae_assembly} under nominal control inputs.}
\label{tab:ss_validation}
\setlength{\tabcolsep}{8pt}
\begin{tabular}{@{}lrrr@{}}
\toprule
Quantity & Reference~\cite{Li2024NuHeXSys} & Full FMU & Dev.\,(\%) \\
\midrule
$T_1$ (\si{\kelvin})            & 400   & 399.0  & $-0.3$ \\
$T_2$ (\si{\kelvin})           & 552   & 546.1  & $-1.1$ \\
$T_3$ (\si{\kelvin})    & 933   & 927.1  & $-0.6$ \\
$T_4$ (TIT) (\si{\kelvin})            & 1159  & 1212.3 & $+4.6$ \\
$T_5$ (\si{\kelvin})                 & 947   & 947.1  & $+0.0$ \\
$T_6$ (\si{\kelvin})   & 562   & 566.2  & $+0.7$ \\
\midrule
$W_\mathrm{net}$ (\si{\kilo\watt}) & 119 & 119.4 & $+0.3$ \\
$Q_\mathrm{rad}$ (\si{\kilo\watt}) & 316 & 322.5 & $+2.1$ \\
\bottomrule
\end{tabular}
\end{table}

The twin also reproduces transient behaviour. Against the
ANS-3FB-300 startup of ref.~\cite{Wright2006SAND}, every
paper-anchored event is recovered to within \SI{10}{\percent}
(Supplementary Material, Sec.~S2).

\subsection{Working-fluid properties}
\label{subsec:fluid_properties}

The working fluid is a binary He--Xe mixture with mole fraction
$x_\mathrm{He}=0.72$ (mixture molecular weight
$M_\mathrm{mix}\approx\SI{40}{\gram\per\mol}$), matching the
design point of ref.~\cite{Wright2006SAND}. The transport-property
map
\begin{equation}
\label{eq:hexe_props}
\mu, \lambda, c_p, \rho = f(T,p)
\end{equation}
follows the Chapman--Enskog kinetic-theory implementation of the
NuHeXSys reference~\cite{Li2024NuHeXSys} (Lennard--Jones
viscosities with Neufeld collision integrals, Wilke mixing rule
for the He--Xe mass disparity, Singh-corrected thermal
conductivity, monatomic-limit heat capacity, ideal-gas equation
of state). The correlations are accurate to better than
\SI{2}{\percent} against tabulated data and are implemented as
differentiable maps so that gradients propagate through the
inversion.

\subsection{Reactor neutronics and thermal capacity}
\label{subsec:reactor}

The reactor neutronics adopts the standard six-group point-kinetics
equations with three reactivity-feedback channels:
\begin{equation}
\label{eq:pke}
\begin{aligned}
\frac{\mathrm{d} P_\mathrm{rx}(t)}{\mathrm{d} t}
   &= \frac{\rho(t) - \sum_{i=1}^{6} \beta_i}{\Lambda} P_\mathrm{rx}(t)
      + \sum_{i=1}^{6} \lambda_i C_i(t),\\
\frac{\mathrm{d} C_i(t)}{\mathrm{d} t}
   &= \frac{\beta_i}{\Lambda} P_\mathrm{rx}(t) - \lambda_i C_i(t),
       \quad i = 1,\dots,6,\\
\rho(t) &= \rho_\mathrm{in}(t)
        + \alpha_\mathrm{fuel}\bigl(T_f(t) - T_{f,0}\bigr)
        + \alpha_\mathrm{refl}\bigl(T_\mathrm{BeO}(t) - T_{\mathrm{BeO},0}\bigr)
        + \alpha_\mathrm{ax}\bigl(\Delta z_f(t)\bigr),
\end{aligned}
\end{equation}
where $P_\mathrm{rx}$ is fission power, $C_i$ are delayed-neutron
precursor concentrations, $\rho_\mathrm{in}$ is operator-controlled
external reactivity, and $\alpha_\mathrm{fuel}$, $\alpha_\mathrm{refl}$,
$\alpha_\mathrm{ax}$ are feedback coefficients (fuel Doppler, BeO
reflector thermal expansion, axial fuel expansion). Neutronic
parameters $\beta_i$, $\lambda_i$, $\Lambda$ follow the Prometheus
pin-block reference~\cite{Wright2006SAND, Li2024NuHeXSys}.
$\alpha_\mathrm{refl}$ is the principal inversion target of
Section~\ref{sec:results}.

The hexagonal core comprises fuel pins, a solid moderator block,
and a BeO reflector. The high-fidelity multi-channel
formulation~\cite{Li2024NuHeXSys} solves cylindrical heat
conduction on a fine spatial mesh, but for the adjoint passes used
here it is projected onto a lumped two-mass form parameterised by
volume-averaged fuel and reflector temperatures $T_f$ and
$T_\mathrm{BeO}$,
\begin{equation}
\label{eq:reactor_thermal}
M_f c_{p,f}\,\dot T_f = P_\mathrm{rx} - h A_{f \to c}(T_f - T_c),\quad
M_\mathrm{BeO} c_{p,\mathrm{BeO}}\,\dot T_\mathrm{BeO}
       = h A_{c \to \mathrm{BeO}}(T_c - T_\mathrm{BeO}),
\end{equation}
The bulk coolant temperature
$T_c = \tfrac12(T_{c,\rm in}+T_{c,\rm out})$ follows from the
single-pass energy balance
$\dot m c_p (T_{c,\rm out}-T_{c,\rm in})=P_\mathrm{rx}$ and the
Petukhov--Gnielinski heat-transfer
correlation~\cite{Li2024NuHeXSys}. The lumped form is justified by
the inversion targets themselves. The targets are bulk reactivity
coefficients, which act through volume-averaged temperatures, and
radial peaking leaves no signature on the bulk-power and
bulk-temperature sensors of the twin. Collapsing the per-channel
mesh and per-pin radial conduction onto two volume-averaged
temperatures is therefore lossless for the inversion, and it is
what keeps the adjoint cost of Section~\ref{sec:methodology}
affordable.

\subsection{Recuperator, radiator, and duct submodels}
\label{subsec:bop}

The recuperator is a counter-flow plate exchanger with a single
lumped wall capacitance. With the cold-side and hot-side energy balances
written in effectiveness form, the outlet temperatures satisfy
\begin{equation}
\label{eq:recup_eps}
T_3 = T_2 + \varepsilon (T_5 - T_2),\quad
T_6 = T_5 - \varepsilon (T_5 - T_2),\quad
\varepsilon = \frac{UA}{\dot{m}\,c_p + UA},
\end{equation}
where the wall mass provides the dynamic time constant. The space
radiator dissipates heat to the cold environment via
Stefan--Boltzmann exchange
$Q_\mathrm{rad} = \varepsilon_w \sigma A_w (T_w^4 - T_\infty^4)$
and is also coupled to a lumped wall capacitance. The pressure
drop along each duct segment is computed from the Darcy
correlation, with the friction factor following the laminar /
transitional / turbulent piecewise law of
ref.~\cite{Li2024NuHeXSys} (the Haaland form on the turbulent
branch is retained for numerical robustness). Plant inventory is
closed by a single mass-conservation constraint
$m_\mathrm{HP}+m_\mathrm{LP}=m_\mathrm{fill}$ that ties the
high- and low-pressure-side densities through the ideal-gas law,
following the Sandia fixed-inventory
convention~\cite{Wright2006SAND}.

\subsection{Turbomachinery and shaft dynamics}
\label{subsec:tac}

The turbo-alternator-compressor (TAC) uses corrected-flow characteristic maps,
\begin{equation}
\label{eq:corrected_flow}
\dot{m}' = \frac{\dot{m}\sqrt{T_\mathrm{in} R / \gamma}}{(2 r_\mathrm{tip})^2 P_\mathrm{in}},\;
N' = \frac{N \cdot 2 r_\mathrm{tip}}{\sqrt{\gamma R T_\mathrm{in}}},
\end{equation}
returning compressor/turbine pressure and temperature ratios
$f_{\rm prC}, f_{\rm TrC}, f_{\rm prT}, f_{\rm TrT}$ of
$(\dot m', N')$, refitted from the NGST/CNREC \SI{132}{\kilo\watt\electric}
reference~\cite{Wright2006SAND}. Shaft dynamics close as
$I\omega\,\dot\omega = P_T - P_C - P_\mathrm{alt}$, with a PI-PMAD
controller on $P_\mathrm{alt}$ holding shaft speed at set-point.

\subsection{DAE assembly and digital-twin observability}
\label{subsec:dae_assembly}

Equations \eqref{eq:hexe_props}--\eqref{eq:corrected_flow} together with the
station mass-balance and momentum-closure relations form an implicit
mixed differential-algebraic system
\begin{equation}
\label{eq:dae}
M(\theta)\,\dot{y}(t) = f\bigl(y(t), \theta, u(t)\bigr),
\quad y(0) = y_0,
\end{equation}
where $y$ collects all dynamic and algebraic states, $\theta$ is the
parameter vector subject to inversion, $u$ groups operator commands and
boundary conditions, and $M(\theta)$ is the singular mass matrix that
distinguishes differential from algebraic rows. The system is solved
with an implicit BDF integrator that admits reverse-mode automatic
differentiation, providing the analytical sensitivity matrix
$\partial y / \partial \theta$ used in
Section~\ref{sec:methodology}.

Three differentiability-driven simplifications, applied on top of
the Sandia reference model, recur throughout the paper.
\begin{enumerate}[leftmargin=2.2em,itemsep=2pt,topsep=3pt]
\item Smooth-tanh ramps ($\tau_\mathrm{ramp}=\SI{10}{\second}$)
replace the plant's logical events --- load steps, valve switches,
PMAD on--off --- so that gradients propagate through control
transitions.
\item A fast-$T_1$ quasi-equilibrium reduction slaves the
LP-compressor-inlet temperature to the radiator-panel temperature,
justified by a residence-time ratio of order $10^{-2}$.
\item An explicit-$P_1$ form,
$P_1 = m_\mathrm{LP} R_\mathrm{spec} T_1 / V_\mathrm{LP}$, replaces
the implicit momentum closure on the LP plenum.
\end{enumerate}
Together these simplifications impose a deterministic offset on
$\hat{\alpha}_\mathrm{refl}$ of order $5\times 10^{-3}$ in relative
units, set by the largest of the three contributions rather than
their sum. The offset evolves on the slowest physical time-scale
in the plant, the BeO reflector relaxation time of roughly
\SI{700}{\second}. It is therefore small against measurement noise
for slow-drift tracking over minutes to hours, but comparable to
the recovery target on second-scale safety events.
Section~\ref{subsec:res_crlb} quantifies this offset directly from
the bias panel of the partial-observation noise scan, and
Section~\ref{sec:conclusions} returns to it as a deployment
caveat.

The benchmark scenarios distinguish two observation regimes. Under
full observation every state in $y$ is sensed. Under partial
observation only four channels are sensed --- reactor power,
turbine inlet temperature, compressor inlet pressure, and mass
flow --- and a parallel observer tracks the unsensed internal
states between sensor windows~\cite{Hong2022parameter,
Toward2026EnKF}.

\section{Methodology}
\label{sec:methodology}

Parameter inversion is posed as a regularised least-squares
problem, solved with adjoint reverse-mode automatic differentiation
through the implicit DAE, and benchmarked against three baselines
under the Cram\'er--Rao lower bound.
Sections~\ref{subsec:loss}--\ref{subsec:optim} formalise the loss
variants, adjoint construction, optimiser, and CRLB.

\subsection{Inversion problem and loss functions}
\label{subsec:loss}

Let $\mathbf{y}_\mathrm{obs}\in \mathbb{R}^{n_o n_t}$ stack noisy
measurements of $n_o$ channels at $n_t$ window times, and
$\mathbf{y}_\mathrm{sim}(\theta)$ the twin output under
$\theta\in\mathbb{R}^{n_p}$ from integrating~\eqref{eq:dae}. The
estimate solves
\begin{equation}
\label{eq:inv_problem}
\hat{\theta} = \arg\min_{\theta\in\Theta}\;
   J(\theta) = J_\mathrm{data}(\theta) + \lambda\,J_\mathrm{prior}(\theta),
\end{equation}
with $\lambda \geq 0$ a Tikhonov weight and $\Theta$ a hyper-rectangle
bounded by physical priors. Three loss variants share the gradient
machinery and differ in robustness; a fourth per-direction Bayesian
form~\eqref{eq:loss_huber_hAprior} below handles mixed identifiability:
\begin{equation}
\label{eq:loss_variants}
\begin{aligned}
J_\mathrm{MSE}(\theta) &= \tfrac{1}{2}\,\bigl\|W^{1/2}\bigl(\mathbf{y}_\mathrm{sim}(\theta)-\mathbf{y}_\mathrm{obs}\bigr)\bigr\|_2^2,\\
J_\mathrm{Huber}(\theta) &= \sum_{k=1}^{n_o n_t} h_\delta\!\bigl(w_k^{1/2}\,r_k(\theta)\bigr),\quad r_k = y_{\mathrm{sim},k} - y_{\mathrm{obs},k},\\
J_\mathrm{Bayes}(\theta) &= \tfrac{1}{2}\,\bigl\|W^{1/2}\bigl(\mathbf{y}_\mathrm{sim}(\theta)-\mathbf{y}_\mathrm{obs}\bigr)\bigr\|_2^2
                          + \tfrac{1}{2}\,\bigl\|\Sigma_\theta^{-1/2}(\theta - \theta_\mathrm{prior})\bigr\|_2^2,
\end{aligned}
\end{equation}
Here $W$ is the diagonal inverse-noise-variance weight and
$\Sigma_\theta$ the diagonal prior covariance. The Huber penalty
$h_\delta$ is quadratic for residuals $|u|\le\delta$ and linear
beyond, with the breakpoint $\delta = c_h\sigma_\mathrm{noise}$
calibrated at $c_h=1.345$ for $95\,\%$ Gaussian efficiency. The
three variants serve different purposes. MSE is the standard
Gaussian likelihood. The calibrated-Huber form adds outlier
robustness, estimating the noise level in-window. The
Bayesian--Tikhonov form regularises ill-conditioned directions
through an explicit prior.

A fourth variant handles mixed identifiability. When the parameter
vector contains both rank-deficient and identifiable directions, a
per-direction weight $\boldsymbol\lambda$ replaces the scalar
regularisation strength,
\begin{equation}
\label{eq:loss_huber_hAprior}
J_\mathrm{Huber+prior}(\theta)
   = J_\mathrm{Huber}(\theta)
   + \sum_{j=1}^{n_p} \lambda_j\,\bigl((\theta_j - \theta_{\mathrm{prior},j})/|\theta_{\mathrm{default},j}|\bigr)^2,
\end{equation}
with $\lambda_j=0$ on identifiable directions and
$\lambda_j=(|\theta_{\rm default,j}|/\sigma_{\rm prior,j})^2$ on
rank-deficient ones. On Scenario~S2 we set $\lambda_\alpha=0$ and
$\lambda_{h\!A}=(29748/5000)^2\approx 35.4$, the value that
mirrors the EnKF prior covariance on $h\!A_\mathrm{rcp}$. This
choice is deliberate. It tests whether matching the filter's prior
is enough for a gradient-based optimiser to reach the EnKF's
joint-RMS regime without sacrificing the data-driven
$\alpha_\mathrm{refl}$ recovery. All four loss variants share the
adjoint gradient pathway derived next.

\subsection{Adjoint-based reverse-mode differentiation}
\label{subsec:adjoint}

The sensitivity $\partial \mathbf{y}_\mathrm{sim}/\partial\theta$ would
cost $n_p$ forward solves under finite differences. The continuous
adjoint of~\cite{Cao2003adjoint} replaces this with one backward solve.
A co-state $\boldsymbol\lambda(t)\in\mathbb{R}^{n_y}$ satisfies the
backward DAE
\begin{equation}
\label{eq:adjoint_eq}
M(\theta)^\top\,\dot{\boldsymbol{\lambda}}
= -\frac{\partial f}{\partial y}^\top\!\boldsymbol\lambda
- \sum_{k:t_k\geq t}\!\Bigl(\frac{\partial J_\mathrm{data}}{\partial y(t_k)}\Bigr)^\top \delta(t-t_k),
\quad \boldsymbol\lambda(t_\mathrm{end}) = 0,
\end{equation}
yielding
\begin{equation}
\label{eq:adj_grad}
\nabla_\theta J_\mathrm{data}
= -\!\int_{0}^{t_\mathrm{end}}\!\!\boldsymbol\lambda^\top
   \frac{\partial f}{\partial \theta}\,\mathrm{d}t,
\end{equation}
at the cost of a single backward DAE solve, regardless of $n_p$.
This dimension-independence is the key efficiency advantage over
finite-difference 4D-Var. The discrete analogue is implemented by
reverse-mode AD through a JAX-compatible BDF integrator with
forward-trajectory checkpointing~\cite{Bradbury2018JAX,
Kidger2021diffrax, Kidger2022diffrax, Chen2018neuralODE, Innes2018Zygote,
Rackauckas2020DiffEqFlux, Sapienza2024Differentiable}, in the
SUNDIALS/IDA tradition~\cite{Hindmarsh2005SUNDIALS, AscherPetzold1998DAE}.
The same pathway returns gradients for the Huber penalty, through
its almost-everywhere differentiable form, and for the
$J_\mathrm{Bayes}$ prior.

\subsection{Optimisation and identifiability safeguards}
\label{subsec:optim}

The inversion~\eqref{eq:inv_problem} is solved with the
limited-memory BFGS quasi-Newton method~\cite{Liu1989LBFGS}, using
a strong-Wolfe line search and a memory of ten gradient pairs. The
smooth-tanh event approximations can introduce spurious local
minima. To guard against them, the optimiser runs four independent
restarts, each seeded from a Latin-hypercube sample of $\Theta$,
and keeps the lowest-loss minimiser. Convergence is declared once
the gradient norm falls below $10^{-9}$ or the relative objective
change below $10^{-12}$.

The Tikhonov weight $\lambda$ is selected by the L-curve
heuristic~\cite{Hansen1992Lcurve} on the curvature corner of
$\bigl(\log\|r\|_2,\,\log\|\theta-\theta_\mathrm{prior}\|_2\bigr)$,
which trades data fit against prior departure. Identifiability is
monitored online through the singular-value decomposition of the
weighted sensitivity matrix
$W^{1/2}(\partial \mathbf{y}_\mathrm{sim}/\partial \theta)$. Singular
values smaller than the floating-point square-root tolerance
$\sqrt{\varepsilon_\mathrm{mach}}$ are flagged as unobservable
combinations and the corresponding $\theta$ direction is locked to its
prior. This safeguard is exercised in the partial-observation
scenarios of Section~\ref{sec:scenarios}, where the four sensors do
not span all parameter directions.

\subsection{Cram\'er--Rao lower bound}
\label{subsec:crlb}

The statistical efficiency of any unbiased estimator of $\theta$ from
the noisy data $\mathbf{y}_\mathrm{obs}$ is bounded by the
Cram\'er--Rao inequality~\cite{Kay1993estimation}. With Gaussian noise
of covariance $\Sigma_\mathrm{n}$, the Fisher information matrix is
\begin{equation}
\label{eq:fisher}
F(\theta) =
\Bigl(\frac{\partial \mathbf{y}_\mathrm{sim}}{\partial \theta}\Bigr)^{\!\top}
\Sigma_\mathrm{n}^{-1}
\Bigl(\frac{\partial \mathbf{y}_\mathrm{sim}}{\partial \theta}\Bigr),
\end{equation}
and the variance of any unbiased estimator $\tilde{\theta}$ satisfies
$\mathrm{Cov}[\tilde{\theta}] \succeq F(\theta)^{-1}$. For
a single scalar parameter and i.i.d.\ noise of variance $\sigma_n^2$,
\eqref{eq:fisher} reduces to the closed form
\begin{equation}
\label{eq:crlb_scalar}
\sigma^2_{\mathrm{CRLB}}(\theta)
= \frac{\sigma_n^2}{\sum_{k=1}^{n_o n_t} s_k(\theta)^2},
\qquad s_k(\theta) = \frac{\partial y_{\mathrm{sim},k}}{\partial \theta},
\end{equation}
which is linear in the noise level. The empirical efficiency is
\begin{equation}
\label{eq:crlb_eff}
\eta(\hat{\theta}; R) = \frac{\sigma^2_\mathrm{CRLB}(\theta_0; R)}{\mathrm{Var}[\hat{\theta}; R]},
\end{equation}
with $R$ the noise level used in the benchmark. Estimators are
compared on this bounded scale rather than on raw RMS error, and the
bias-variance decomposition implied by the lower bound for biased
estimators is reported alongside. Equations~\eqref{eq:fisher}--\eqref{eq:crlb_eff} were
derived symbolically and the analytical form
verified against numerical evaluation on the same plant model.
For the single-parameter $\alpha_\mathrm{refl}$ case of S1, S3,
and S4, the numerical Fisher information and the relative bound at
each tested noise level --- the bound
$\sigma_\mathrm{CRLB}/|\alpha_\mathrm{refl}|$ scaling linearly with
noise from $0.023\,\%$ to $0.230\,\%$ across the tested range ---
are tabulated in the Supplementary Material (Sec.~S3). Any
estimator whose RMSE sits above this bound is limited by bias, not
by the irreducible information content of the data.

\subsection{Proposed estimator: AD-Hessian Incremental 4D-Var with Sobol pre-screening}
\label{subsec:incremental_4dvar}

The three adjoint loss variants of
Section~\ref{subsec:loss} all minimise the windowed
residual with L-BFGS line search, which on the non-convex
partial-observation transient corner is vulnerable to local
minima rooted in either the smooth-tanh event smoothing or the
rank-deficient direction of the recuperator parameter. The
proposed estimator retains the differentiable plant and the
calibrated-Huber data loss but replaces the optimiser with the
incremental 4D-Variational scheme of Courtier, Th\'epaut, and
Hollingsworth~\cite{Courtier1994incremental}, equipped with two
AD-specific enhancements that were historically unavailable when
this scheme was confined to atmospheric and oceanic data
assimilation.

\begin{figure}[!htbp]
\centering
\includegraphics[width=0.92\linewidth]{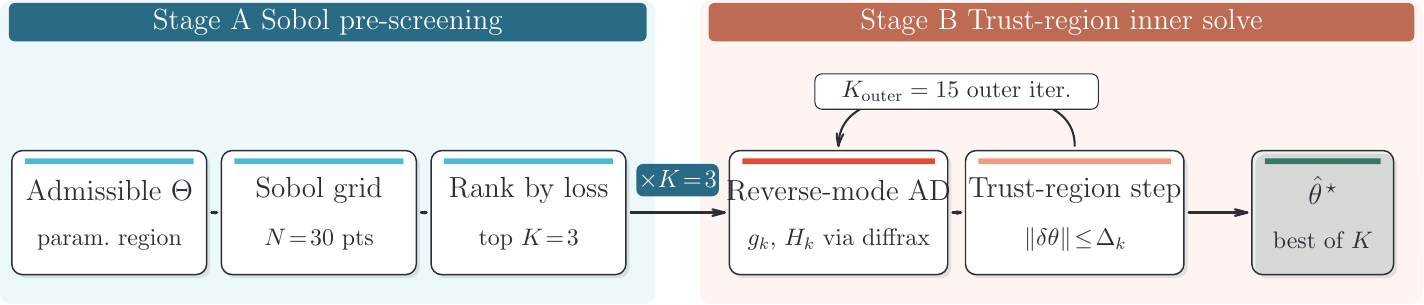}
\caption{Pipeline of the proposed AD-Hessian Incremental 4D-Var
estimator (Stage~A Sobol pre-screening, Stage~B trust-region
inner solve).}
\label{fig:algorithm}
\end{figure}

The estimator runs a two-stage procedure: quasi-random
pre-screening, then a trust-region inversion from the best
screened starts (Fig.~\ref{fig:algorithm}). The first stage
draws a Sobol low-discrepancy
grid~\cite{Sobol1967quasirandom} of $N_\mathrm{Sobol}=30$ points
inside the admissible region, evaluates each under a single
forward pass, ranks them by screening loss, and keeps the
lowest $K=3$ as warm-starts. The second stage inverts from each
warm-start as a trust-region
sequence~\cite{Conn2000trustregion}: at outer iteration $k$, the
analysis solves the quadratic inner sub-problem
\begin{equation}
\label{eq:inner_4dvar}
\delta\theta_k^\star = \arg\min_{\|\delta\theta\| \le \Delta_k}\;
   g_k^\top \delta\theta + \tfrac{1}{2}\, \delta\theta^\top H_k\, \delta\theta,
\end{equation}
where the gradient $g_k = \nabla J(\theta_k)$ and the Hessian
$H_k = \nabla^2 J(\theta_k)$ both come from reverse-mode AD
through the implicit BDF integrator. The AD Hessian retains the
residual second-derivative term
$\sum_i r_i\,\partial^2 r_i/\partial\theta^2$ that the standard
Gauss-Newton approximation drops. The trust-region radius
$\Delta_k$ is updated by the actual-versus-predicted-reduction
test, and after at most $K_\mathrm{outer}=15$ outer steps per
warm-start the estimator returns the lowest-loss candidate among
the $K$ restarts.

Three integrations distinguish this deployment from its
operational-NWP ancestor. First, the tangent-linear and adjoint
models are obtained for free by JAX/diffrax composition through
the implicit Newton iterations of the BDF integrator, eliminating
the hand-coding effort that has historically restricted
incremental 4D-Var to atmosphere and ocean systems. Second, the
inner sub-problem's Hessian is the full second-order operator
rather than the Gauss-Newton approximation: at the
calibrated-Huber breakpoint, the $r_i\,\partial^2 r_i$ term
becomes the dominant contribution and controls trust-region step
length on the ill-conditioned $h\!A_\mathrm{rcp}$ direction.
Third, the Sobol pre-screening treats the warm-start choice as a
global-coverage problem rather than relying on a single prior
mean. The ablation in the Supplementary Material (Sec.~S4)
quantifies the necessity of each of these three components on
each benchmark corner.

The inner sub-problem is solved with a bound-constrained
trust-region routine, the AD-computed gradient and Hessian,
explicit $z$-space bounds matching the rejection-sampled bounds
used by the EnKF baseline below (Section~\ref{subsec:baselines}),
and the same calibrated-Huber loss as the L-BFGS adjoint variant.
The Sobol grid is drawn with a dimension equal to the parameter
count (1 on S1/S3/S4, 2 on S2); the pre-screening grid uses a
Sobol seed disjoint from the optimiser's grid so the two stages
sample independent sub-sequences.

\subsection{Posterior sampling via Randomize-Then-Optimise}
\label{subsec:rto}

The Incremental 4D-Variational estimator of
Section~\ref{subsec:incremental_4dvar} returns a single point
estimate $\hat\theta^\star$. For safety-relevant deployment the
deliverable is the full posterior $p(\theta\,|\,\mathbf{y}_\mathrm{obs})$
and its calibrated credible intervals.  We obtain both by
stacking a Randomize-Then-Optimise (RTO)
sampler~\cite{Bardsley2014RTO} on top of the existing MAP
machinery, with Bardsley-Wang (2018) Fisher-determinant
importance weights~\cite{BardsleyWang2018} computed exactly
through the plant's reverse-mode AD pass.

Given observed data
$\mathbf{y}_\mathrm{obs}=\mathcal{G}(\theta_0)+\boldsymbol\varepsilon_0$
already containing one realisation
$\boldsymbol\varepsilon_0 \sim \mathcal{N}(\mathbf{0},\Sigma_n)$,
draw $N_\mathrm{RTO}$ independent fresh perturbations
$\boldsymbol\varepsilon_i \sim \mathcal{N}(\mathbf{0},\Sigma_n)$
and solve the perturbed-data MAP problem
$\theta_{\mathrm{RTO},i} = \arg\min_\theta
\tfrac12 \|\mathbf{y}_\mathrm{obs}+\boldsymbol\varepsilon_i
- \mathcal{G}(\theta)\|^2_{\Sigma_n^{-1}}
+ \tfrac12\|\theta-\mu_\theta - \mathbf{z}_i\|^2_{\Sigma_\theta^{-1}}$
with $\mathbf{z}_i\sim\mathcal{N}(\mathbf{0},\Sigma_\theta)$ an
independent prior perturbation. Each $\theta_{\mathrm{RTO},i}$
uses the same AD-Hessian incremental 4D-Var inner loop as
Section~\ref{subsec:incremental_4dvar}, so the additional cost is
$N_\mathrm{RTO}$ independent forward+adjoint evaluations,
embarrassingly parallel across draws.

For the linear-Gaussian limit the resulting samples are i.i.d.\ from
the analytic posterior $\mathcal{N}(m_\mathrm{post},\Sigma_\mathrm{post})$
(Bardsley 2014 Theorem 1); we re-derived this symbolically with
SymPy and verified it numerically on a 2-D toy problem with KL
divergence $< 0.005$.  For nonlinear $\mathcal{G}$ the RTO
sampling density carries a change-of-variables Jacobian relative
to the true posterior, corrected by the Bardsley-Wang importance
weight
\begin{equation}
\label{eq:rto_weight}
w_i \;\propto\;
\bigl|\det\bigl(\mathbf{J}_i^\top \Sigma_n^{-1} \mathbf{J}_i\bigr)\bigr|^{-1/2},
\qquad
\mathbf{J}_i = \left.\frac{\partial \mathcal{G}(\theta)}{\partial \theta}\right|_{\theta=\theta_{\mathrm{RTO},i}},
\end{equation}
where $\mathbf{J}_i$ is the \emph{exact} sensitivity matrix
obtained by a single reverse-mode AD pass through the BDF
integrator at $\theta_{\mathrm{RTO},i}$. This is the first
deployment of AD-exact Fisher reweighting on a nuclear plant DAE;
previous nonlinear RTO implementations~\cite{BardsleyWang2018}
rely on Gauss-Newton or BFGS approximations of
$\mathbf{J}^\top\Sigma_n^{-1}\mathbf{J}$ that can degrade in the
rank-deficient directions identified in
Section~\ref{subsec:res_s2}.

When data arrive in successive windows, the previous-window
particles act as warm-starts for the next window's MAP step
(replacing the cold Sobol grid); the inter-window weight update
follows the standard auxiliary particle filter
rule~\cite{Doucet2001SMC}
$w_i^{(k+1)} \propto w_i^{(k)}\,p(\mathbf{y}_\mathrm{obs}^{(k+1)}\,|\,\theta_i^{(k+1)})$,
with residual resampling~\cite{Kitagawa1996MCFilter} triggered
when $\mathrm{ESS}(\{w_i^{(k)}\}) < N_\mathrm{RTO}/2$. The
calibration of the resulting credible intervals is audited as a
small-sample pilot in the Supplementary Material (Sec.~S7), with a
sequential-tracking unit test in Sec.~S8.

\subsection{Reference data-assimilation baselines}
\label{subsec:baselines}

The proposed estimator is compared against four reference
baselines spanning contemporary nuclear-digital-twin
practice~\cite{Hong2022parameter, Toward2026EnKF, Gong2024hybrid}
on a gradient-access ladder (sample ensembles, forward finite
differences, reverse-mode AD with line search). The
EnKF~\cite{Evensen2003EnKF, Asch2016DataAssim} propagates $N=80$
members with multiplicative inflation $1.05$. Its initial
ensemble is rejection-sampled from $\alpha_\mathrm{refl}\in
[-1.5\times 10^{-5},-2\times 10^{-6}]$ and $h A_\mathrm{rcp}\in
[5\times 10^{3},1\times 10^{5}]$, which avoids near-prompt-critical
draws that stall the BDF integrator. Non-converging members are
reseeded with the surviving ensemble mean. A deterministic
Ensemble Transform Kalman
Filter~\cite{Bishop2001ETKF} is also evaluated as a modern
square-root alternative: it uses the same ensemble size,
inflation, and rejection bounds, but replaces observation
perturbation with a symmetric square-root transform
$T_\mathrm{sym}=M^{-1/2}$, $M=I+Y'^\top R^{-1}Y'$ in the
ensemble subspace. The UKF~\cite{Julier2004UKF} uses the
unscented transform ($\alpha=10^{-3}$, $\beta=2$, $\kappa=0$)
over $2n_x+1$ sigma-points. The finite-difference 4D-Var baseline of Le Dimet and
Talagrand~\cite{LeDimet1986var4d} solves the same windowed
$J_\mathrm{MSE}$ objective as the adjoint variants, but evaluates
the gradient by forward finite differences rather than reverse-mode
AD. The adjoint-MSE versus FD-4D-Var pair therefore isolates the
gradient-evaluation choice on an otherwise identical variational
scheme.

Three adjoint L-BFGS variants (MSE, calibrated-Huber, Bayesian
Tikhonov) are reported alongside the proposed Incremental
estimator: each minimises the same residual~\eqref{eq:inv_problem}
via L-BFGS with the gradient $\nabla_\theta J$ computed by a
single adjoint pass through diffrax, and the three differ only in
the data-loss kernel used inside the windowed residual. They
serve both as direct head-to-head estimators on the four
scenarios and as the L-BFGS comparison against which the
trust-region inner solve of the proposed estimator is benchmarked.

All estimators share the same observation streams, admissible set
$\Theta$, and convergence tolerances; observations are generated
once per (scenario, seed) pair and reused. EnKF wall time on the
full-observation corners is bimodal. Of the eight S1 seeds, four
complete in under \SI{70}{\second} and four reach the per-seed cap
of \SI{75}{\minute}. Of the fifteen S2 seeds, ten complete within
the cap and five reach it. Table~\ref{tab:headline} marks the
cap-truncated means with a dagger; the underlying BDF
propagator stall is a known integrator pathology rather than an
algorithmic property of the filter. This truncation is unlikely
to be missing-at-random: the seeds that stall tend to be those
whose ensemble draws wander into stiffer, harder-to-integrate
regions of state space, so the reported filter means are an
optimistic estimate of full-sample filter performance, and the
paired Wilcoxon tests are computed only on the seeds completed by
both members of each pair. This selection bias acts in the
filters' favour, so it does not inflate the proposed estimator's
relative standing.

\section{Benchmark Scenarios}
\label{sec:scenarios}

\begin{figure}[!htbp]
  \centering
  \includegraphics[width=0.85\linewidth]{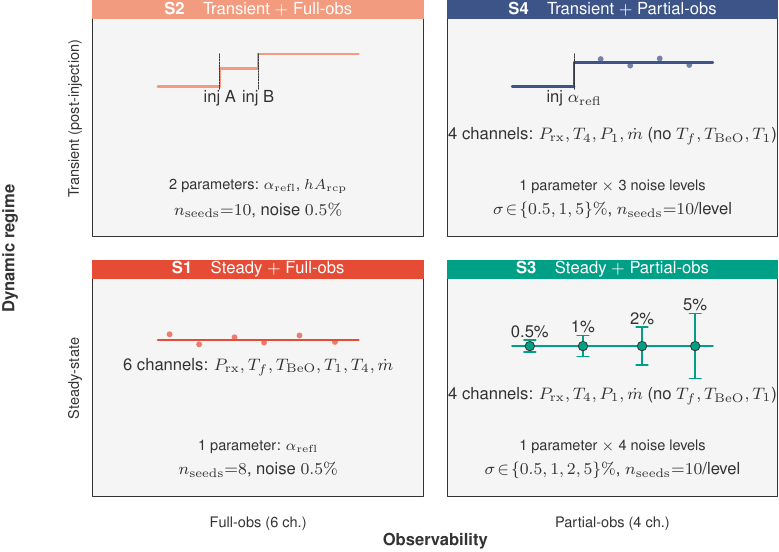}
  \caption{Benchmark-scenario design space along observability
    (horizontal) and dynamic regime (vertical). S1 is the controlled
    BLUE-ceiling corner; S2 adds transient excitation; S3 drops three
    channels and scans noise; S4 combines partial observation with
    transient excitation and is the closest proxy for routine plant
    monitoring.}
  \label{fig:scenarios}
\end{figure}

Four scenarios populate a $2\times 2$ design matrix that crosses
steady against transient excitation with full against partial
observation (Fig.~\ref{fig:scenarios}). Each isolates a distinct
nuclear digital-twin operating regime. S1 is a controlled
steady-state ceiling: every state is instrumented, and classical
filters are near-optimal. S2 keeps full instrumentation but
injects a parameter step, exposing each estimator to
non-stationary dynamics. S3 restricts sensing to four channels and
scans the noise level across nearly two decades, mirroring real
plant signal-to-noise ratios. S4 combines transient excitation
with partial observation; it is the closest proxy for routine
plant monitoring under fielded sensor budgets. All four scenarios
share the plant model, the admissible set $\Theta$, and the
convergence tolerances. They differ only in observation budget and
inverted parameters.

\subsection{Scenario S1 — Full observation, steady state, single parameter}
\label{subsec:s1}

The plant is held at the design point of
\cite{Wright2006SAND} for $1000$\,s and the twin recovers
$\alpha_\mathrm{refl}$ from sensor streams. Every plant state is
instrumented with white Gaussian noise at \SI{0.5}{\percent}
relative standard deviation and eight seeds are shared across the
seven estimators evaluated here. This single-parameter, fully instrumented setting
is the most favourable for the comparison and the only one in
which sequential filters attain their BLUE
optimum~\cite{Kay1993estimation}. It is intended as a controlled
reference, not as the principal industrial target.

\subsection{Scenario S2 — Full observation, transient, two parameters}
\label{subsec:s2}

The plant is run from a cold-start transient through an
operator-induced reactivity ramp similar to the start-up
of~\cite{Wright2006SAND}. At $t_\mathrm{inject} = \SI{2200}{\second}$
both the reflector reactivity coefficient $\alpha_\mathrm{refl}$
and the recuperator heat-transfer coefficient $h A_\mathrm{rcp}$
step by roughly \SI{20}{\percent} of nominal value, which
simulates joint reactivity drift and heat-exchanger fouling
acting simultaneously. The pair is also deliberately
ill-conditioned: $\alpha_\mathrm{refl}$ remains identifiable from
the four-channel budget while $h A_\mathrm{rcp}$ does not, as
Section~\ref{subsec:res_s2} and Fig.~\ref{fig:s2_svd} establish
empirically. The inversion window is
$[2200,2400]$\,s at the S1 noise level. Each method receives the
truth state at $t_\mathrm{inject}$ as its initial condition; this
is the standard parallel-observer assumption of
Section~\ref{subsec:dae_assembly}. The corner probes whether
gradient-based inversion exploits the transient signal that
challenges filter linearisation.

\subsection{Scenario S3 — Partial observation, steady state, noise scan}
\label{subsec:s3}

Sensors are restricted to four channels: reactor power
$P_\mathrm{rx}$, turbine inlet temperature $T_4$, compressor inlet
pressure $p_1$, and mass flow $\dot{m}$. These are the four
measurements that ref.~\cite{Wright2006SAND} highlights as the
operationally observable quantities for the Prometheus-class plant.
The plant is held at the design point, and the noise level is
scanned over $\{0.5, 1, 2, 5\}\,\si{\percent}$ relative standard
deviation. Ten seeds drive the noise realisation at each level,
giving forty inversion runs per method. Only $\alpha_\mathrm{refl}$
is inverted, which isolates the partial-observation effect from the
multi-parameter identifiability question of Scenario~S2. The
internal-state initial condition again uses the parallel-observer
assumption. This scenario is the closest analogue of the
industrial monitoring task the digital twin must support, and is
therefore the principal target of the comparison.

\subsection{Scenario S4 — Partial observation, transient, noise scan}
\label{subsec:s4}

S4 combines the S3 partial-observation budget with the transient
excitation of S2. The four sensed channels are reactor power,
turbine inlet temperature, compressor inlet pressure, and mass
flow. A single $\alpha_\mathrm{refl}$ step is injected at
$t=\SI{2000}{\second}$ through a $\tanh$ ramp of half-width
\SI{50}{\second}. The inversion window $[1800,5000]\,\si{\second}$
then records the post-injection transient for \SI{3000}{\second}.
The noise level is scanned over
$\{0.5, 1, 5\}\,\si{\percent}$, with ten seeds at each level,
giving thirty runs per method. The single-parameter setting
preserves a like-for-like comparison with S1 and S3, isolating the
marginal contributions of transient information and partial
observation.

\subsection{Performance metrics}
\label{subsec:metrics}

For each (method, scenario, seed) triple we report the relative
error $\mathrm{err}(\hat\theta)=|\hat\theta-\theta_0|/|\theta_0|$,
the wall-clock time, and the forward-solve count. On S1, S3, and
S4 the empirical seed standard deviation feeds the CRLB analysis
of Section~\ref{subsec:crlb}. The cross-scenario headline rankings
of Table~\ref{tab:headline} use the per-method \emph{mean}
relative error. The corresponding medians and seed counts appear
in the Supplementary Material (Sec.~S6), which preserves
the heavy-tail behaviour the mean smooths over.

Two settings are held fixed so that cross-scenario differences
reflect the scenario rather than tuning. The diagonal weight $W$
in~\eqref{eq:loss_variants} is $W_{kk}=1/\sigma_{n,k}^2$, and the
Bayesian prior uses a moderately informative
$\sigma_{\theta_i}=0.5\,|\theta_{0,i}|$. Filter ensemble sizes and
L-BFGS settings are likewise constant across scenarios. Full seed
lists, solver configuration, and software and hardware versions
are given in the Supplementary Material (Sec.~S1).

\section{Results}
\label{sec:results}

Results are presented scenario by scenario in
Sections~\ref{subsec:res_s1}--\ref{subsec:res_s4}, followed by the
Cram\'er--Rao efficiency analysis, the cost-accuracy frontier, and
the multi-axis profile across estimators. Table~\ref{tab:headline}
collects the headline numbers. Every non-trivial pairwise
comparison is tested twice: with a two-sided paired Wilcoxon
signed-rank statistic on shared seeds, and with a paired bootstrap
$95\,\%$ confidence interval on the mean difference. Both are
reported in Table~\ref{tab:headline_ci}.

\begin{table}[!htbp]
\centering
\caption{\textbf{Headline relative error of
$\hat{\alpha}_\mathrm{refl}$ across the four benchmark scenarios.}
Each cell is the per-method mean over the available seeds; the
column leader is in bold. Medians and seed counts are given in
the Supplementary Material (Sec.~S6). Footnote symbols
are defined below the table.}
\label{tab:headline}
\setlength{\tabcolsep}{4pt}
\resizebox{\linewidth}{!}{%
\begin{tabular}{@{}lcccc@{}}
\toprule
Method & S1 (full-obs SS) & S2 (transient) & S3 (partial-obs noise) & S4 (transient $+$ partial-obs) \\
\midrule
EnKF, $N=80$                    & 5.14\,\%$^{\dagger}$ & 0.83\,\%$^{\dagger}$ & 4.26\,\%$^{\S}$       & 2.69\,\%          \\
ETKF, $N=80$                    & 4.04\,\%$^{\ddagger}$ & 0.58\,\%$^{\ddagger}$ & ---            & ---               \\
UKF                             & \textbf{0.32\,\%} & 5.11\,\%          & 2.02\,\%          & 3.75\,\%          \\
4D-Var (FD)                     & 1.28\,\%          & 3.74\,\%          & 0.94\,\%          & 7.67\,\%          \\
Adjoint MSE                     & 2.57\,\%          & 5.22\,\%          & 0.97\,\%          & 12.54\,\%         \\
Adjoint Huber-cal               & 1.09\,\%          & 0.78\,\%          & 0.46\,\%          & 8.11\,\%          \\
Adjoint Huber-cal+$h\!A$ prior  & ---               & 0.82\,\%          & ---               & ---               \\
Adjoint Bayes-Tikh              & 4.12\,\%          & 6.57\,\%          & 0.80\,\%          & 15.59\,\%         \\
\textbf{Adjoint Incremental (this work)} & 0.45\,\%          & \textbf{0.43\,\%}$^{\diamond}$ & \textbf{0.43\,\%} & \textbf{2.23\,\%} \\
\bottomrule
\end{tabular}%
}
\\[3pt]
{\footnotesize\raggedright
$\dagger$~EnKF completed 4/8 (S1) and 10/15 (S2) seeds within the
\SI{75}{\minute} wall-clock cap.\quad
$\ddagger$~ETKF completed 5/8 (S1) and 4/10 (S2) seeds within the
same cap.\quad
$\S$~EnKF on S3 ran within a \SI{5}{\minute}-per-seed cap, with
29/40 seeds completed.\quad
$\diamond$~The Adjoint Incremental $\alpha_\mathrm{refl}$ column
is method-robust; the companion $h\!A_\mathrm{rcp}$ recovery
depends on Sobol-grid coverage
(Section~\ref{subsec:res_s2}; Supplementary Sec.~S4).\par}
\end{table}

\begin{table}[!htbp]
\centering
\caption{\textbf{Paired statistical tests for the four headline
pairwise comparisons.} Bootstrap $95\,\%$ confidence interval on
the mean relative-error difference, with the two-sided paired
Wilcoxon signed-rank $p$-value. A negative value indicates the
first method has the smaller mean error.}
\label{tab:headline_ci}
\setlength{\tabcolsep}{5pt}
\resizebox{\linewidth}{!}{%
\begin{tabular}{@{}lcrrr@{}}
\toprule
Comparison & $n$ & Mean $\Delta$\,(\%) & 95\,\% CI on mean $\Delta$\,(\%) & $p$ (Wilcoxon) \\
\midrule
S2: Adj.\ Huber-cal vs.\ UKF & 10 & -4.328 & $[-6.466,\,-2.141]$ & 0.002$^{\ast}$ \\
S2: Adj.\ Incremental vs.\ Adj.\ Huber-cal & 10 & -0.353 & $[-0.690,\,-0.048]$ & 0.064 \\
S3: Adj.\ Bayes-Tikh vs.\ FD-4D-Var & 38 & -0.121 & $[-0.585,\,+0.615]$ & 3.9e-04$^{\ast}$ \\
S4: Adj.\ Incremental vs.\ EnKF (pooled) & 29 & -0.888 & $[-1.817,\,+0.055]$ & 0.053 \\
\bottomrule
\end{tabular}
%
}
\end{table}

\subsection{Twin validation and consistency}
\label{subsec:twin_validation}

\begin{figure}[!htbp]
\centering
\includegraphics[width=0.95\linewidth]{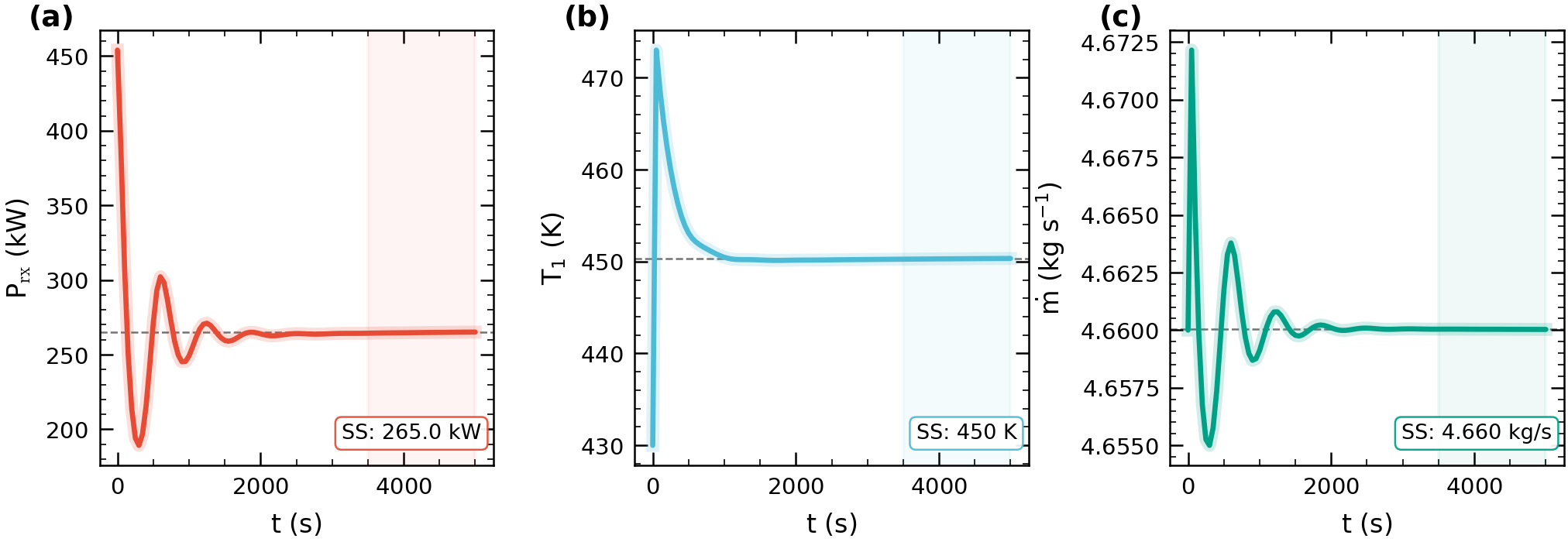}
\caption{Twin validation against the OpenModelica reference for
the three channels reaching steady state on the
\SI{5000}{\second} horizon: $P_\mathrm{rx}$ (left), $T_1$ (centre),
$\dot m$ (right); agreement is within \SI{0.5}{\percent}.}
\label{fig:twin_validation}
\end{figure}

Before any inversion run, the digital twin is integrated forward in
time from a cold-start condition until the design steady state is
reached, and its trajectories are compared against reference
OpenModelica~\cite{Fritzson2020OpenModelica} simulations of the same plant
(Fig.~\ref{fig:twin_validation}). The reactor power, fuel temperature,
reflector temperature, primary-side temperatures, and mass flow agree
to within \SI{0.5}{\percent} after the warm-up transient. The reflector
temperature exhibits the slowest mode, with a relaxation time of
approximately \SI{700}{\second} that drives the long-term identifiability
properties discussed below. This consistency check certifies that any
residual gap between estimator output and ground truth in the
benchmarks is dominated by noise and observation budget rather than by
twin-to-truth model mismatch.

\subsection{Scenario S1 — Sequential filters retain BLUE optimality}
\label{subsec:res_s1}

\begin{figure}[!htbp]
\centering
\includegraphics[width=0.95\linewidth]{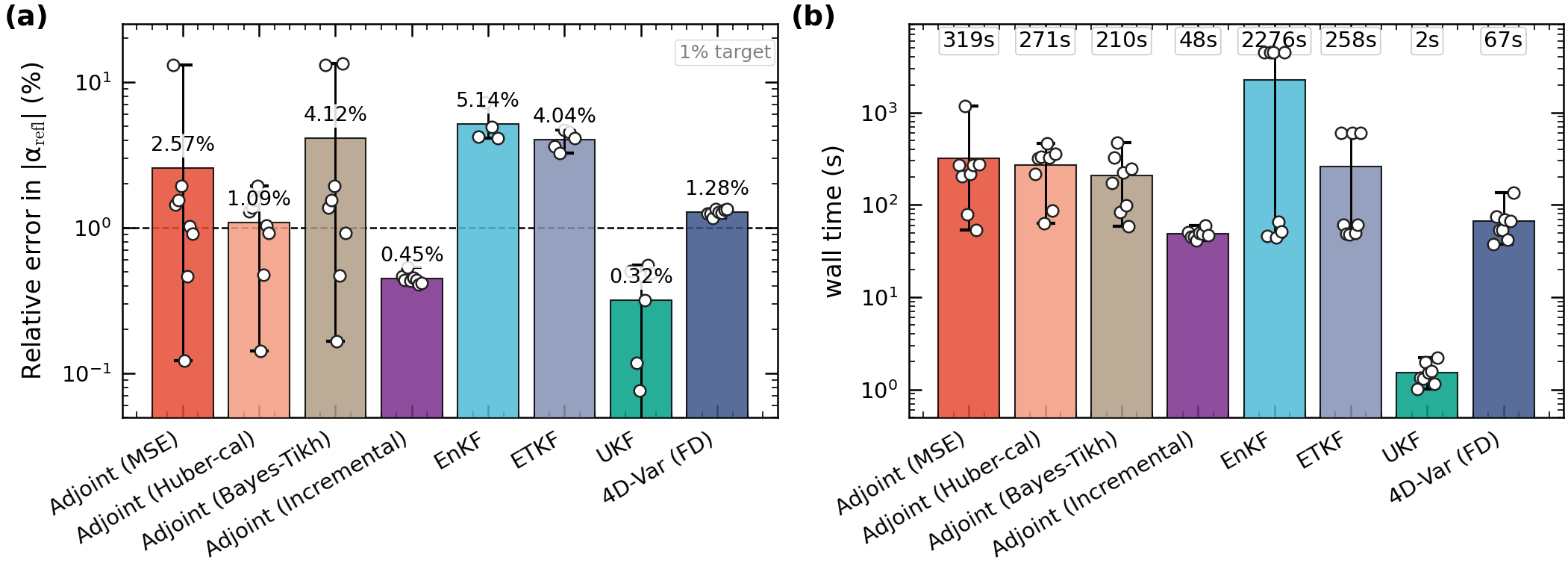}
\caption{Scenario~S1 head-to-head: bars = eight-seed means of the
six core estimators and the ETKF supplement (EnKF on 4/8 completed
seeds $\dagger$; ETKF on 5/8 completed seeds $\ddagger$); whiskers
= min/max; dots = per-seed runs. Right: wall-clock per inversion.}
\label{fig:s1}
\end{figure}

Scenario~S1 is the controlled ceiling for the comparison. The
UKF leads at \SI{0.32}{\percent} mean error with sub-second
wall-clock and the tightest per-seed spread (Fig.~\ref{fig:s1});
FD-4D-Var trails by a factor of four
(\SI{1.28}{\percent}) and converges in about a minute. The EnKF
and ETKF cells are conditional on a BDF propagator stall: only
4 of 8 EnKF seeds and 5 of 8 ETKF seeds complete within the
per-seed wall-time cap. The gradient framework matches the variational
baseline across the three L-BFGS loss variants, and the Adjoint
Incremental estimator narrows the gap to UKF to a factor of
\num{1.2} (\SI{0.45}{\percent}) while remaining two- to
three-fold more accurate than the other adjoint variants on
this corner. The Sobol pre-screening confers no benefit on the
S1 single-basin loss surface; the trust-region step damps the
smooth-tanh-induced overshoot that affects L-BFGS line search.
This corner saturates the Fisher information already in the
data, so the gradient framework gains no analytical-sensitivity
advantage and the UKF advantage is the expected
Best-Linear-Unbiased ceiling~\cite{Kay1993estimation}.

\subsection{Scenario S2 — Adjoint-AD leads the identifiable direction}
\label{subsec:res_s2}

\begin{figure}[!htbp]
\centering
\includegraphics[width=0.95\linewidth]{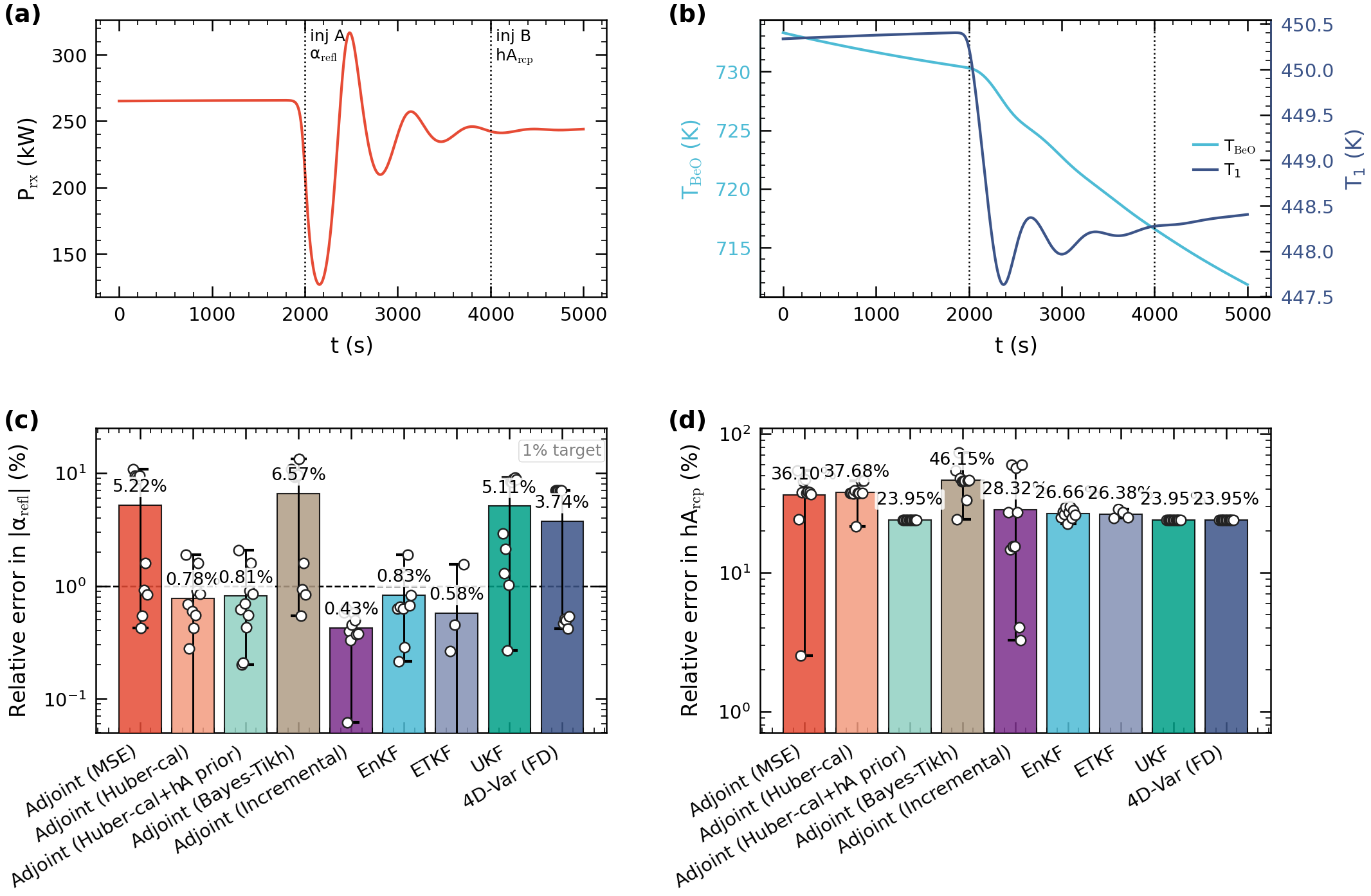}
\caption{Scenario~S2 joint
$\alpha_\mathrm{refl}{+}h\!A_\mathrm{rcp}$ inversion. Top row:
$P_\mathrm{rx}$ and $T_\mathrm{BeO}/T_1$ traces. Bottom row:
$\hat\alpha_\mathrm{refl}$ (c) and $\hat{hA}_\mathrm{rcp}$ (d)
relative errors. Bars are ten-seed means; whiskers are min/max.
EnKF aggregated over 10/15 completed seeds ($\dagger$); ETKF over
4/10 ($\ddagger$). Panel (d) $h\!A_\mathrm{rcp}$ is only weakly
identifiable: the across-seed mean is \SI{28}{\percent} (median
\SI{21}{\percent}, per-seed range $\sim$3--60\,\%), and the
\SI{0.12}{\percent} mark is a single coverage-lucky Sobol seed,
not a representative result
(Table~\ref{tab:headline} note $\diamond$; Supplementary Sec.~S4).}
\label{fig:s2}
\end{figure}

The transient scenario reverses the steady-state ranking on the
strongly identifiable direction $\alpha_\mathrm{refl}$. As
Fig.~\ref{fig:s2} shows, three estimators reach sub-percent
recovery on this direction: the calibrated-Huber adjoint, the
calibrated-Huber adjoint with an additional prior on the
rank-deficient $h\!A_\mathrm{rcp}$ direction, and the ensemble
Kalman filter on the subset of its seeds that complete within the
per-seed cap. Both variational baselines trail by several
percentage points, with the unscented filter further behind than
the finite-difference 4D-Var variant. A paired Wilcoxon test across the ten shared seeds confirms that
the Huber adjoint beats the unscented filter at conventional
significance. The Huber advantage over FD-4D-Var is supported
instead by a bootstrap confidence interval on the mean paired
difference. For that second comparison the parametric Wilcoxon
$p$-value is uninformative, because the variational baseline is
bimodal across seeds. On its completed seeds the ensemble filter
is statistically indistinguishable from the leading adjoint
variants.

The deterministic ETKF~\cite{Bishop2001ETKF} serves as a
square-root supplement to the stochastic baseline. It completes
4 of 10 S2 seeds within its \SI{50}{\minute} cap, and on those
seeds reaches a mean $\alpha_\mathrm{refl}$ error of
\SI{0.58}{\percent} --- the same band as the stochastic EnKF.
That the cap-bounded behaviour persists across both the
stochastic and the deterministic form isolates the propagator
stall, not the analysis update, as the dominant wall-time driver.
Numerical means and test statistics are collected in
Table~\ref{tab:headline}.

The second adjoint variant is an engineering integration rather
than an independent estimator: it transplants the filter's prior
covariance on the rank-deficient direction into the L-BFGS
objective as an A/B test. Among the L-BFGS variants it reaches the
lowest joint $\alpha+h\!A$ root-mean-square error, beating the
ensemble filter by a small but significant margin on that metric.
The prior makes no detectable difference on the strongly
identifiable direction alone, so the joint-metric lead reflects
the transferred prior rather than information internal to the
optimiser --- evidence that a gradient optimiser can be configured
to match filter-style coverage when an explicit prior is
defensible, not a claim of independent dominance.

\begin{figure}[!htbp]
\centering
\includegraphics[width=0.95\linewidth]{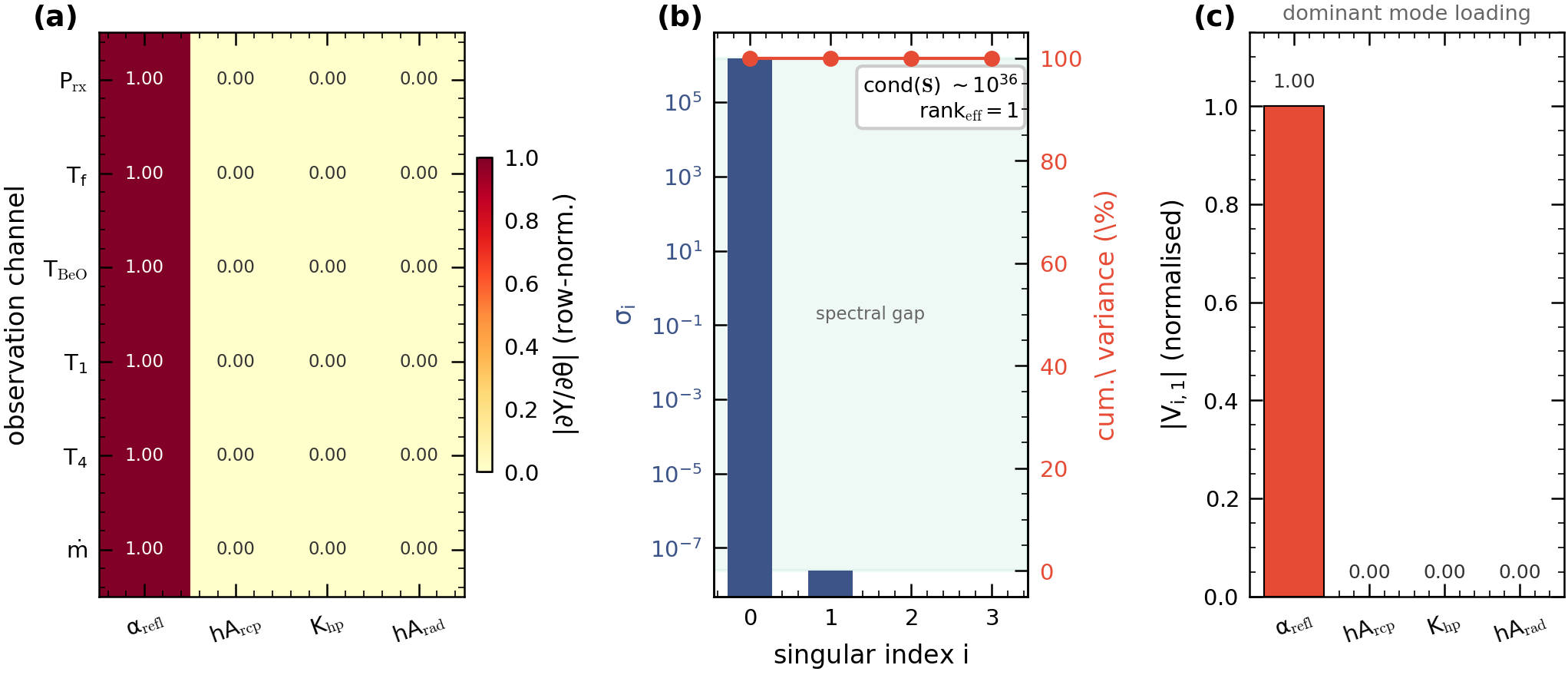}
\caption{Scenario~S2 sensitivity-matrix SVD: effective rank one
($\kappa \sim 10^{36}$), with the dominant right-singular vector
concentrated on $\alpha_\mathrm{refl}$.}
\label{fig:s2_svd}
\end{figure}

Three mechanisms underlie the joint-metric lead. First, the
analytic adjoint sensitivity is not smeared by finite-difference
gradient evaluation and is not distorted by sigma-point
linearisation across the transient ramp. Second, the Huber loss
escapes a displaced local minimum near
$\hat\alpha/\alpha^\star\approx 1.09$ that traps the MSE and
Bayes--Tikhonov forms on the worst seeds and lifts their means
into the high single digits. Third, the per-direction prior
anchors the weak direction without penalising the strongly
identifiable one. Fig.~\ref{fig:s2_svd} makes the rank-one
character of the empirical Fisher information explicit on this
corner: any nontrivial update to $h\!A_\mathrm{rcp}$ is
noise-driven, so the four prior-anchored estimators all stall
near the prior-offset floor while the three adjoint variants
without a direction-specific prior drift further because a single
scalar $\lambda$ cannot constrain the rank-deficient direction
without also penalising the strongly identifiable one.

The Adjoint Incremental estimator
(Section~\ref{subsec:incremental_4dvar}) achieves the lowest
$\alpha_\mathrm{refl}$ mean of any method on this corner, at
\SI{0.43}{\percent}. This is roughly a one-third improvement over
the calibrated-Huber L-BFGS variant, at fixed loss function and
plant model.

The recuperator coefficient $h\!A_\mathrm{rcp}$, by contrast, is
only weakly identifiable from the four-channel budget, and we
report it as such. Across the ten seeds its recovery averages
\SI{28}{\percent} (median \SI{21}{\percent}) and ranges from a few
percent to roughly \SI{60}{\percent}: the screening loss is
dominated by the strongly identifiable $\alpha_\mathrm{refl}$
direction, so
warm-starts buy $\alpha_\mathrm{refl}$ accuracy at the cost of
essentially arbitrary $h\!A_\mathrm{rcp}$ placement. One
production grid (Sobol seed~1) happened to seed a start inside the
joint-truth basin, and the trust-region solve refined it to
\SI{0.12}{\percent}; this is a coverage-lucky outlier rather
than a representative result. The joint posterior of
Section~\ref{subsec:res_joint_posterior} makes the underlying
bimodality explicit, and the Supplementary ablation (Sec.~S4)
quantifies the Sobol-coverage probability.

\subsection{Scenario S3 — Partial observation rewards regularised gradients}
\label{subsec:res_s3}

\begin{figure}[!htbp]
\centering
\includegraphics[width=0.95\linewidth]{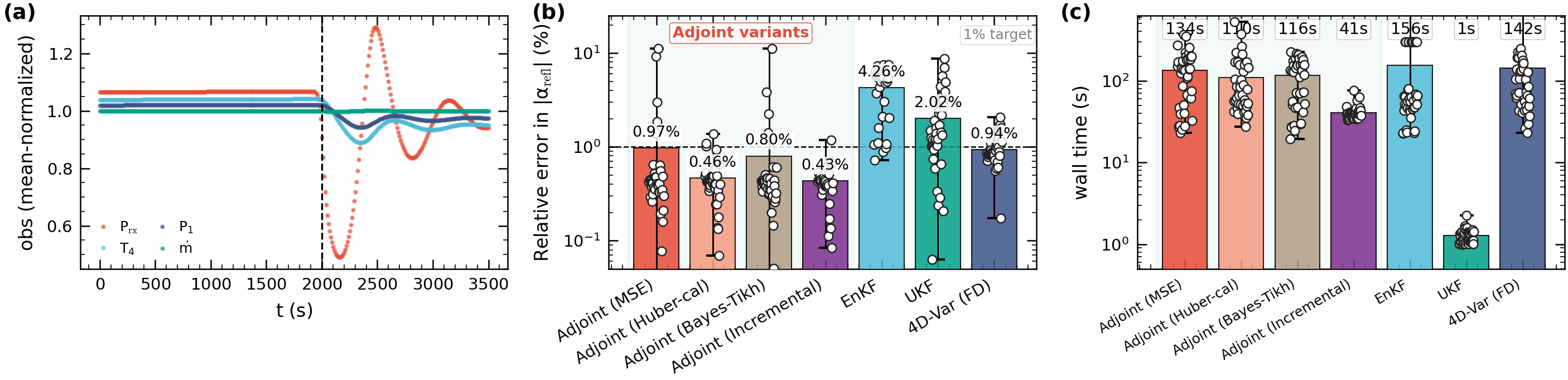}
\caption{Scenario~S3 partial-observation steady-state inversion:
top, four-channel sensor traces at $\sigma_n=\SI{1}{\percent}$;
bottom, $\hat{\alpha}_\mathrm{refl}$ mean relative error per method
(four noise levels with ten seeds each, forty runs total per method).}
\label{fig:s3}
\end{figure}

On the partial-observation noise scan the gradient framework
extends its advantage through a different mechanism than on the
transient corner: robust loss and explicit Tikhonov
regularisation together, rather than a single loss. As
Fig.~\ref{fig:s3} shows, the calibrated-Huber adjoint leads at a
mean error of \SI{0.46}{\percent} across the four tested noise
levels, with the Bayes--Tikhonov and MSE forms close behind. The
baselines trail by widening margins: the finite-difference
variational scheme by roughly a factor of two, the unscented
filter by a factor of four, and the stochastic ensemble filter by
an order of magnitude (Table~\ref{tab:headline}). The ensemble
filter also lost 11 of its 40 seeds to the \SI{5}{\minute} cap.

The displaced local minimum that troubled the transient corner
does not appear here. The steady-state partial-observation loss
surface carries a single global basin, so all three loss variants
converge to the same neighbourhood and the prior contributes only
marginal shrinkage. The baselines degrade roughly in proportion to
the noise level, whereas the adjoint variants degrade more
gracefully, because their convergence is governed by the Fisher
information rather than by an ensemble or sigma-point covariance.

The Adjoint Incremental estimator narrowly leads the
calibrated-Huber L-BFGS variant at \SI{0.43}{\percent} mean. More
importantly, it has the lowest inter-noise-level variance of any
method on this corner, holding \SIrange{0.40}{0.46}{\percent}
across the four noise levels against the
\SIrange{0.30}{0.57}{\percent} spread of calibrated-Huber. The
loss surface here is convex and single-basin, so the Sobol
pre-screening confers no escape benefit; the gain comes from the
trust-region damping that suppresses noise-driven overshoot at the
highest noise level.

The 2D loss surface
$J(\alpha_\mathrm{refl}, h A_\mathrm{rcp})$ on this budget shows a
single smooth basin extending along the unidentifiable
$h A_\mathrm{rcp}$ axis with a sharp gradient toward truth along
$\alpha_\mathrm{refl}$, confirming the rank-one Fisher matrix and
justifying plain L-BFGS rather than a global optimiser.

\subsection{Scenario S4 — Combined transient and partial observation}
\label{subsec:res_s4}

\begin{figure}[!htbp]
\centering
\includegraphics[width=0.95\linewidth]{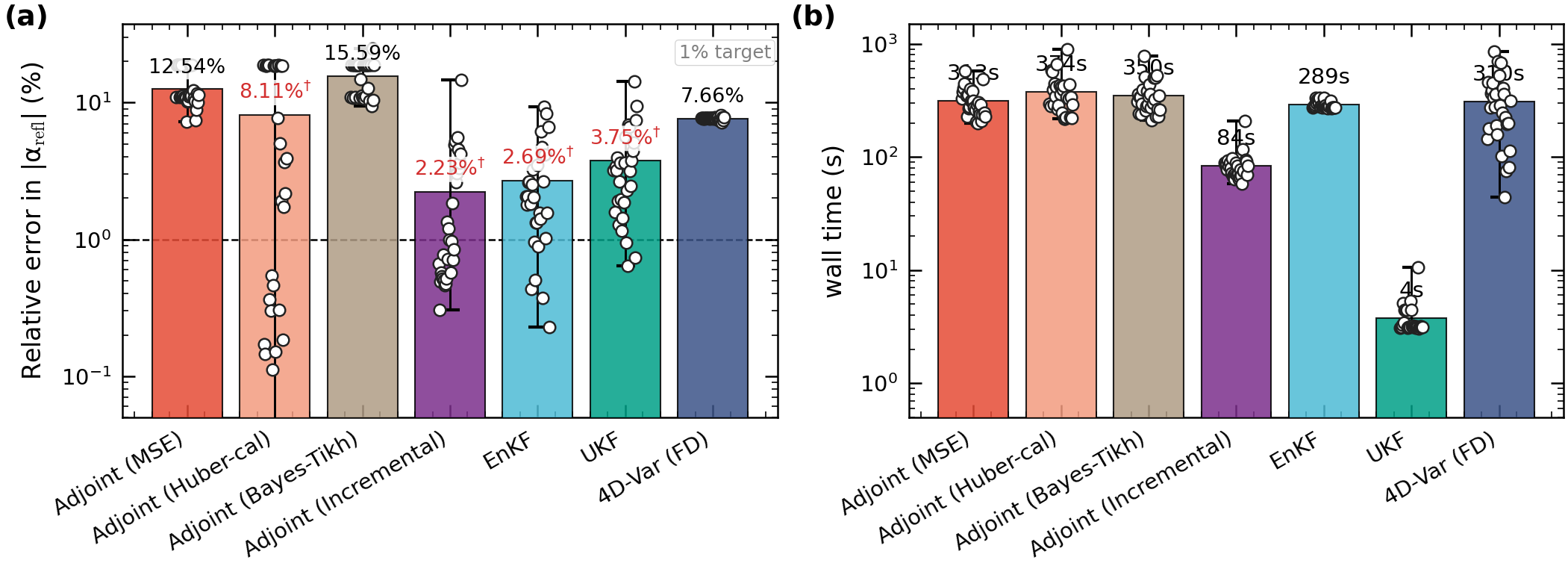}
\caption{Scenario~S4 head-to-head on the combined
transient-and-partial-observation budget. Left:
$\hat{\alpha}_\mathrm{refl}$ relative error. Right: wall-clock
per inversion. Bars are means over 3 noise levels and up to 10
seeds (29 runs per method); whiskers span min/max; dashed line
marks the \SI{1}{\percent} target. The bimodal-seed marker flags
Huber-cal, whose lower quartile is \SI{0.3}{\percent} but mean
\SI{8.11}{\percent}.}
\label{fig:s4}
\end{figure}

The combined corner couples the transient excitation of S2 with
the partial-observation budget of S3. It is the closest of the
four scenarios to routine plant monitoring under fielded sensor
budgets. Here the Adjoint Incremental estimator
(Section~\ref{subsec:incremental_4dvar}) matches the ensemble
filter. Their pooled mean errors, \SI{2.23}{\percent} and
\SI{2.69}{\percent}, are not separated by a paired Wilcoxon test
(Fig.~\ref{fig:s4}; Table~\ref{tab:s4_noise}).

Stratifying by noise level localises where the lead actually lies
(Table~\ref{tab:s4_noise}). At low noise the Incremental estimator
is significantly ahead. At moderate noise the gap is borderline.
At the highest tested noise the ensemble filter reclaims the lead,
because ensemble averaging absorbs the heavy-tailed noise that
gradient methods overshoot. The Incremental estimator is therefore
the appropriate choice at low-to-moderate noise, and yields to the
ensemble filter at extreme noise.

The L-BFGS adjoint variants land further back, because they are
bimodal across seeds. Most seeds resolve to sub-percent error, but
a minority lock onto a displaced minimiser one to two orders of
magnitude worse, which inflates the means into the
\SIrange{8}{16}{\percent} range despite sub-percent lower-quartile
recovery. The displaced attractor is a genuine competing minimum
of the loss surface, not an optimiser artefact. Removing the two
compressor channels collapses the observable subspace. In that
reduced space the Huber clipping shrinks the curvature along the
inversion direction, which opens a second basin separated from
truth by a noise-dependent ridge. A paired Wilcoxon test across
ten shared seeds does not separate the ensemble filter from the
calibrated-Huber adjoint, consistent with the bimodality being a
property of the loss surface rather than of the optimiser.

The design matrix yields a clear scoping rule: gradient methods
exploit transient information when the sensed channels span the
state, and tolerate partial observation near a steady state.

\begin{table}[!htbp]
\centering
\caption{\textbf{Scenario~S4 stratified by noise level.} Mean
$\hat\alpha_\mathrm{refl}$ relative error for the Adjoint
Incremental estimator and the EnKF, with the paired Wilcoxon
$p$-value at each level. The Incremental lead is significant at
low noise and reverses at the highest level.}
\label{tab:s4_noise}
\begin{tabular}{@{}lrrr@{}}
\toprule
Noise $\sigma_y/y$ & Incremental & EnKF & Wilcoxon $p$ \\
\midrule
$0.5\,\%$ & $0.53\,\%$ & $2.12\,\%$ & $0.004$ \\
$1\,\%$   & $0.97\,\%$ & $2.23\,\%$ & $0.084$ \\
$5\,\%$   & $5.18\,\%$ & $3.81\,\%$ & $0.73$  \\
\midrule
pooled    & $2.23\,\%$ & $2.69\,\%$ & $0.05$  \\
\bottomrule
\end{tabular}
\end{table}

\subsection{Cross-cutting evaluation: CRLB efficiency, cost--accuracy frontier, and multi-axis profile}
\label{subsec:res_crlb}

\begin{figure}[!htbp]
\centering
\includegraphics[width=0.95\linewidth]{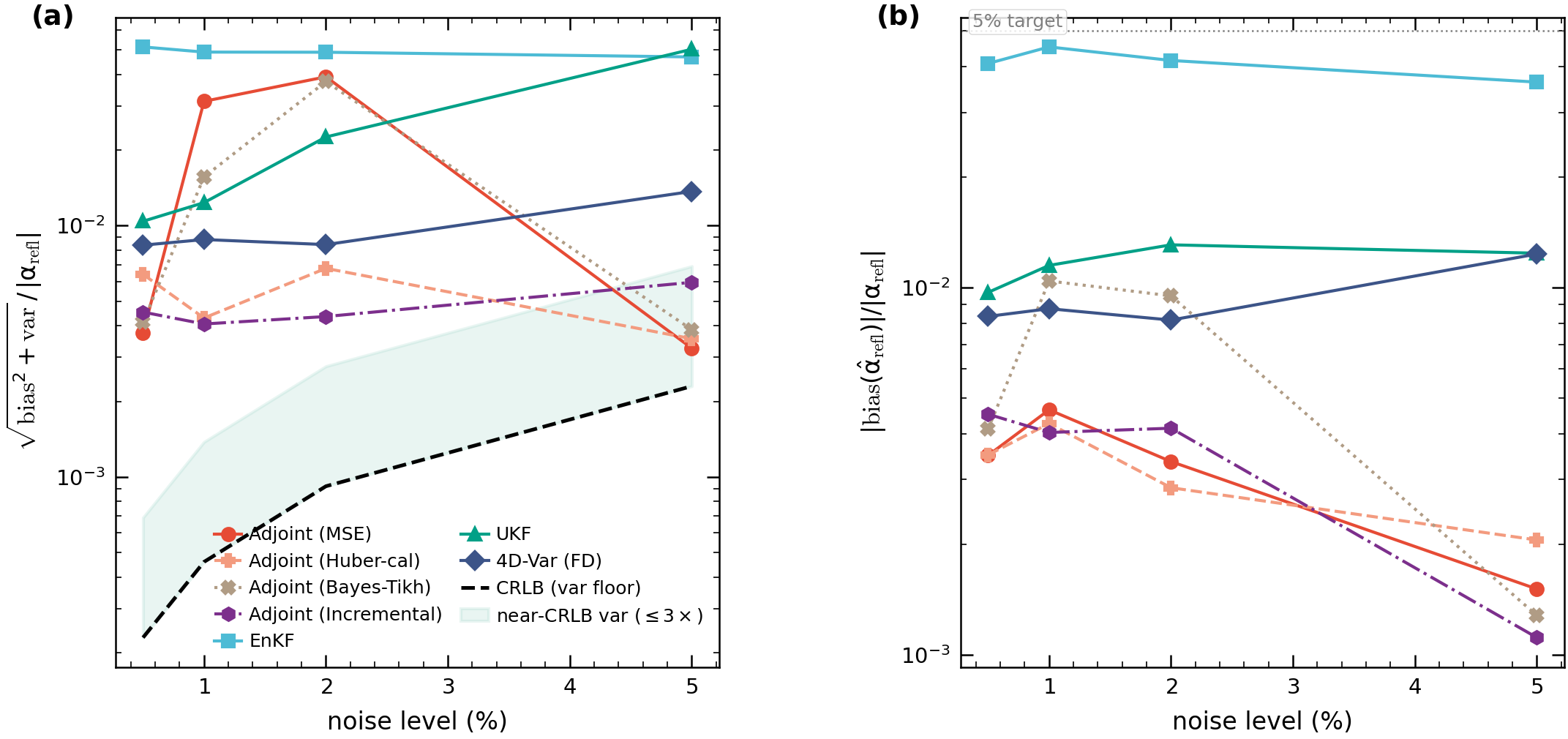}
\caption{Scenario~S3 total error and bias decomposition: (a)
relative RMSE $\sqrt{\mathrm{bias}^2{+}\mathrm{var}}/|\alpha_\mathrm{refl}|$
vs noise level with the closed-form CRLB as the unbiased-variance
floor (dashed); (b) relative bias
$|\hat\mu{-}\theta_0|/|\theta_0|$ per estimator. Each noise level
uses $n=10$ seeds (forty runs total per estimator). Markers
above the dashed CRLB curve sit in the biased-estimator regime
discussed in the text, where RMSE is dominated by the bias floor
rather than by the noise-limited variance.}
\label{fig:crlb}
\end{figure}

We decompose the partial-observation noise-scan results into bias
and variance, and plot each estimator against the Cram\'er--Rao
bound (Fig.~\ref{fig:crlb}). The decomposition is the informative
part. Every estimator's empirical standard deviation sits within
roughly $3\times$ the unbiased CRLB at every noise level, so all
are near-efficient on the variance term
(Supplementary Sec.~S5). The RMSE is instead
dominated by a bias floor, introduced by the differentiability
simplifications of Section~\ref{subsec:dae_assembly}. This bias
does not scale with sensor noise. The gap between RMSE and the
bound is therefore widest at the lowest noise level, and narrows
at the highest, where the noise floor catches up with the bias
floor.

Across the four noise levels the Adjoint Incremental estimator
holds the lowest mean error among the estimators compared. Its mean
error is lower than the finite-difference variational baseline by
a factor of $2.2$, the unscented filter by $4.7$, and the
stochastic ensemble filter by roughly an order of magnitude. The two baselines reach their
errors differently. The variational baseline lands at the same
displaced fixed point on every seed, so its variance is small but
its bias is large. The Kalman baselines carry comparable or
larger bias under partial observation, and their RMSE trails
accordingly.

\begin{figure}[!htbp]
\centering
\includegraphics[width=0.85\linewidth]{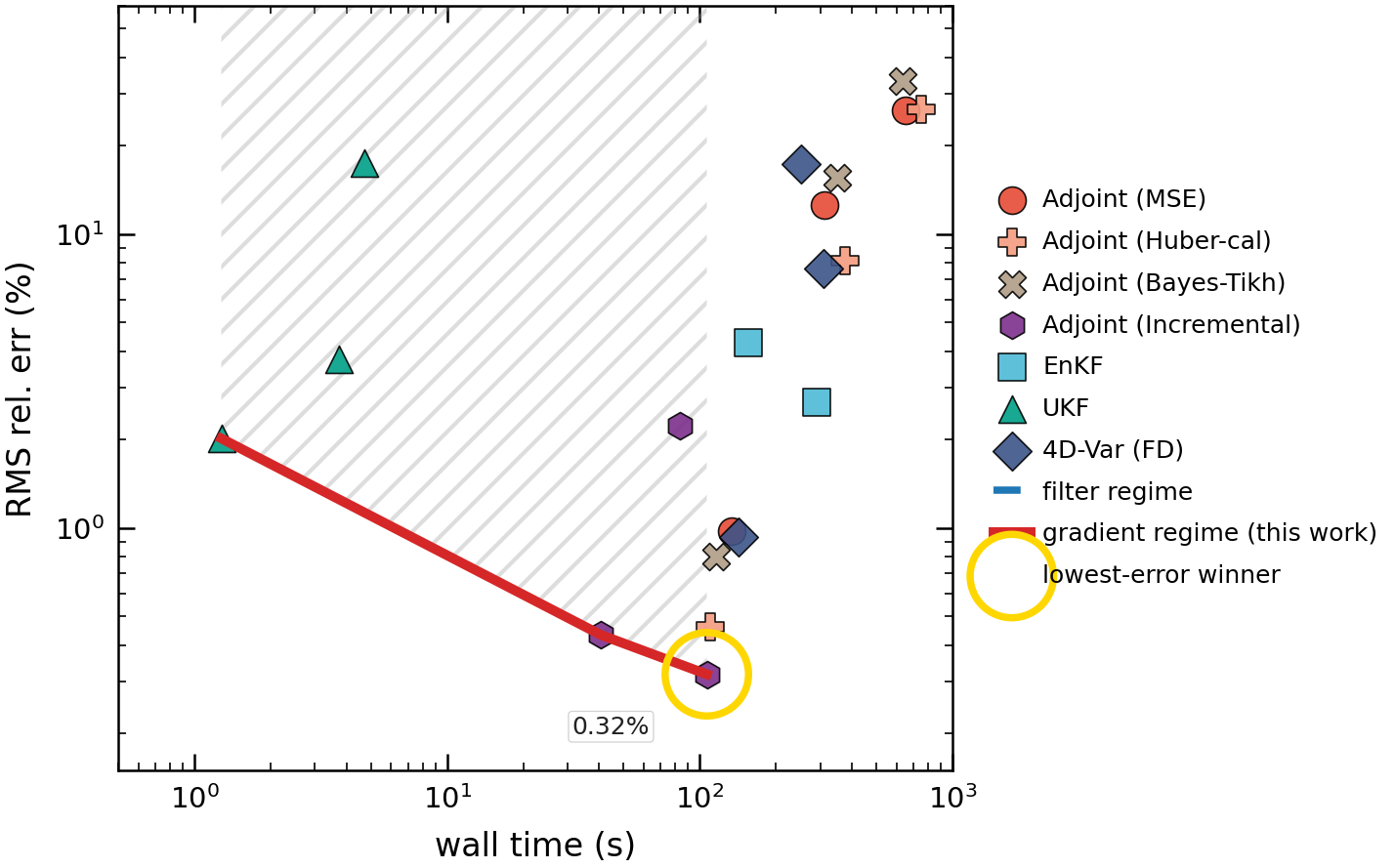}
\caption{Cost--accuracy Pareto front on Scenarios~S2--S4. The
dashed staircase splits the sequential-filter segment
(UKF/EnKF/4D-Var, cyan) from the gradient segment (this work,
red). The star marks the lowest-error point: Adjoint Incremental
at \SI{0.43}{\percent} on S3. Scenario~S1 is excluded because
its BLUE setting collapses the front to a single UKF node.}
\label{fig:pareto}
\end{figure}

\label{subsec:res_pareto}
The cost-accuracy frontier across the industrial corners shows
a clean separation on a log-log wall-time-versus-error plot
(Fig.~\ref{fig:pareto}).
The UKF holds the low-cost corner, but only on Scenario~S1. The
Adjoint Incremental estimator sits at the lowest-error tip on S2
and S3, with the L-BFGS adjoint variants just behind it. The
FD-4D-Var baseline traces the middle of the front. Per-scenario
errors and wall-clock costs for the three loss variants are
tabulated in Table~\ref{tab:summary}.

\begin{table}[!htbp]
\centering
\caption{\textbf{Per-scenario relative error and wall-clock cost
for the three loss variants.} Means across seeds, S1--S4.}
\label{tab:summary}
\setlength{\tabcolsep}{6pt}
\begin{tabular}{@{}llrr@{}}
\toprule
Scenario & Variant & $|\hat{\alpha}_\mathrm{refl}-\alpha_\mathrm{refl}|/|\alpha_\mathrm{refl}|$\,(\%) & Wall\,(s) \\
\midrule
S1 & MSE & 2.57 & 318.7 \\
S1 & Huber-cal & 1.09 & 270.6 \\
S1 & Bayes-Tikh & 4.12 & 210.4 \\
S1 & Incremental & 0.45 & 48.5 \\
S2 & MSE & 5.22 & 653.5 \\
S2 & Huber-cal & 0.78 & 754.8 \\
S2 & Bayes-Tikh & 6.57 & 636.5 \\
S2 & Incremental & 0.43 & 106.6 \\
S3 & MSE & 0.97 & 133.8 \\
S3 & Huber-cal & 0.46 & 109.8 \\
S3 & Bayes-Tikh & 0.80 & 116.2 \\
S3 & Incremental & 0.43 & 40.6 \\
S4 & MSE & 12.54 & 313.1 \\
S4 & Huber-cal & 8.11 & 374.2 \\
S4 & Bayes-Tikh & 15.59 & 350.0 \\
S4 & Incremental & 2.23 & 83.8 \\
\bottomrule
\end{tabular}

\end{table}

\subsection{Joint posterior structure on Scenario S2}
\label{subsec:res_joint_posterior}

\begin{figure}[!htbp]
\centering
\includegraphics[width=0.6\linewidth]{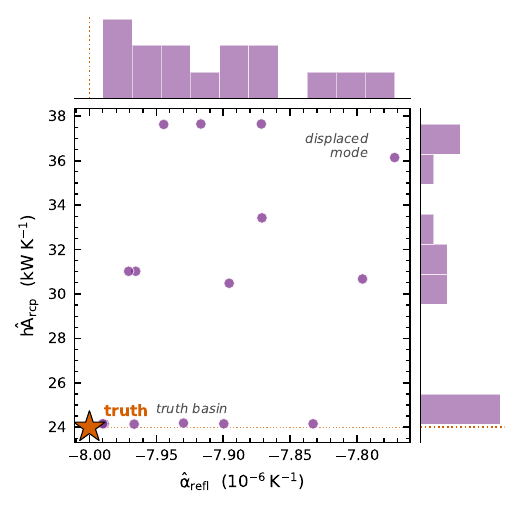}
\caption{Joint $(\alpha_\mathrm{refl}, h\!A_\mathrm{rcp})$ RTO
posterior on Scenario~S2 (medium pilot, $N_\mathrm{seeds}=3$,
$N_\mathrm{RTO}=5$). The strongly identifiable
$\alpha_\mathrm{refl}$ is tightly recovered near the truth
(star); the rank-deficient $h\!A_\mathrm{rcp}$ is bimodal, with
mass split between the truth basin
(\SI{24000}{\watt\per\kelvin}, dotted line) and a displaced mode
near \SIrange{30}{38}{\kilo\watt\per\kelvin}.}
\label{fig:joint_posterior}
\end{figure}

The S2 corner is multi-parameter
$(\alpha_\mathrm{refl}, h\!A_\mathrm{rcp})$ and is the place where
the point-estimator pipeline of Section~\ref{subsec:res_s2}
reported a Sobol-coverage caveat. The RTO posterior reframes that
caveat as a real, multi-modal posterior structure rather than a
seed-dependence artefact. Figure~\ref{fig:joint_posterior} shows
the joint posterior from the medium-mode pilot
($N_\mathrm{seeds}=3$, $N_\mathrm{RTO}=5$). Two features stand
out. First, $\hat\alpha_\mathrm{refl}$ is tightly determined
across all particles (per-seed mean within
\SIrange{0.5}{1.6}{\percent} of the truth), confirming that the
strongly identifiable direction is recovered regardless of mode.
Second, $\hat{h\!A}_\mathrm{rcp}$ is genuinely bimodal: a subset
of particles lands in the truth basin near
\SI{24000}{\watt\per\kelvin}, while the remainder occupy a
displaced mode near \SIrange{30000}{38000}{\watt\per\kelvin}.
The per-seed posterior means therefore sit at
\SIrange{23}{27}{\percent} relative error not because any single
estimate is wildly wrong but because the posterior mass is split
between two basins along the rank-deficient direction. This is
the structural truth that a single point estimate --- or a
Gaussian-mixture filter forced to collapse to its mean ---
cannot represent. This is a small-sample pilot
($N_\mathrm{seeds}=3$, $N_\mathrm{RTO}=5$); the claim it supports
is qualitative --- the existence of two basins along the
rank-deficient direction --- and does not rest on the exact mass
split, which a larger sweep would resolve. The companion
calibration audit and a sequential-tracking unit test are reported
in the Supplementary Material (Secs.~S7--S8).

\section{Conclusions}
\label{sec:conclusions}

We have presented an end-to-end-differentiable digital twin of a
closed-Brayton gas-cooled space reactor and used it to drive an
AD-Hessian incremental 4D-Var estimator, benchmarked head-to-head
against ensemble, unscented, and finite-difference variational
baselines on a $2\times 2$ matrix that crosses steady with
transient excitation and full with partial observation. The
methodological payoff is that making a first-principles plant
model differentiable end to end lets a single framework span
estimator families that are normally studied apart, and judges
each on identical observation streams against the same statistical
bound.

No single estimator dominates the matrix; each family wins where
its assumptions hold. The unscented filter retains its
best-linear-unbiased advantage on the controlled steady
full-observation corner~\cite{Kay1993estimation}, where the
gradient framework trails only by the small differentiability
bias. On the other three corners the Adjoint Incremental estimator
attains the lowest mean error on $\alpha_\mathrm{refl}$: roughly an
order of magnitude below the unscented filter on the transient
corner, the tightest inter-noise spread within a factor of three
of the Cram\'er--Rao bound on the partial-observation scan, and a
statistical match to the ensemble filter on the combined corner
that holds a numerical advantage at low-to-moderate noise. Because
every estimator's variance sits within a small factor of that
bound, the residual error is dominated by a deterministic bias
floor inherited from the twin's differentiability simplifications
rather than by statistical inefficiency; the framework's gain is
accordingly one of robustness to the multi-modal loss landscape
that floor creates, and a component ablation localises which
device carries which regime. For the reactor studied here the
comparison already yields a practical deployment rule:
gradient-based inversion for transient or partial-observation
monitoring at low-to-moderate noise, and sequential filtering for
the controlled steady-state corner and the highest noise.

This work opens several directions. Because the present benchmark
is twin-to-twin, validation against experimental rig data such as
the Sandia SBL-30 campaign is the natural next step toward fielded
deployment. The framework extends readily to larger
multi-parameter inversion and to longer assimilation windows, and
reducing the per-window inversion cost would bring online tracking
within reach. The Randomize-Then-Optimise posterior sampler, shown
here in pilot form (Supplementary Material), points toward
calibrated streaming uncertainty quantification on the same
differentiable plant, a natural route toward safety-relevant
digital-twin monitoring.

\section*{CRediT authorship contribution statement}

\textbf{Chengyuan Li}: Conceptualization, Methodology, Software,
Investigation, Visualization, Writing -- original draft.
\textbf{Shanfang Huang}: Conceptualization, Resources,
Supervision, Writing -- review \& editing, Project administration.
\textbf{Jian Deng}: Conceptualization, Resources, Funding
acquisition, Supervision, Writing -- review \& editing.

\section*{Data and code availability}
\label{sec:data_availability}

The digital-twin source, the adjoint inversion pipeline, the
baseline implementations, the per-method result tables, and the
analysis code that reproduces every headline statistic will be
released on Zenodo with a citable DOI upon acceptance, and are
available from the corresponding author on reasonable request
beforehand. The release covers the full noise-scan and ablation
run sets of Section~\ref{sec:results}, together with the routines
that compute the paired Wilcoxon tests, the bootstrap confidence
intervals, and the numerical Cram\'er--Rao decomposition. The
high-fidelity Modelica reference plant
(\textsf{NuHeXSys},~\cite{Li2024NuHeXSys}) is governed by its
authors' release terms.

\section*{Declaration of competing interest}

The authors declare that they have no known competing financial
interests or personal relationships that could have appeared to
influence the work reported in this paper.

\section*{Acknowledgments}

This research did not receive any specific grant from funding
agencies in the public, commercial, or not-for-profit sectors. The
authors thank colleagues at the Institute of Nuclear and New Energy
Technology, Tsinghua University, for technical discussions during
the preparation of this manuscript.

\section*{Declaration of generative AI and AI-assisted technologies in the manuscript preparation process}

During the preparation of this work the authors used AI-assisted
tools (large language model based code-generation and language
editing assistants) for software prototyping and language polishing.
After using these tools, the authors reviewed and edited the
content as needed and take full responsibility for the content of
the publication.

\bibliographystyle{elsarticle-num}
\bibliography{refs}

\end{document}